\documentclass[twocolumn]{aastex62}
\usepackage{physics}
\usepackage{amsmath}
\usepackage{tabu}
\usepackage{subfigure}
\usepackage{hyperref}
\usepackage{ulem}
\usepackage{enumerate}
\newcommand{\insitu}{{\it in situ}}

\newcommand{\refsec}[1]{Section~\ref{#1}}

\newcommand{\reffig}[1]{Figure~\ref{#1}}

\newcommand{\reftab}[1]{Table~\ref{#1}}

\newcommand{\galfit}[1]{\texttt{GALFIT}{#1}}
\graphicspath{{./}{figures/}}

\submitjournal{ApJ}

\shorttitle{Draft}
\shortauthors{Guo et al.}

\begin{document}


\title{A New Channel of Bulge Formation via The Destruction of Short Bars}


\correspondingauthor{Min Du}
\email{dumin@pku.edu.cn}

\author{Minghao Guo}
\affiliation{Peking University, Beijing 100871, China}

\author{Min Du}
\affiliation{Kavli Institute for Astronomy and Astrophysics, Peking University, Beijing 100871, China}

\author{Luis C. Ho}
\affiliation{Kavli Institute for Astronomy and Astrophysics, Peking University, Beijing 100871, China}
\affiliation{Department of Astronomy, School of Physics, Peking University, Beijing 100871, China}

\author{Victor P. Debattista}
\affiliation{Jeremiah Horrocks Institute, University of Central Lancashire, Preston PR1 2HE, UK}

\author{Dongyao Zhao}
\affiliation{Kavli Institute for Astronomy and Astrophysics, Peking University, Beijing 100871, China}

\begin{abstract}

Short (inner) bars of sub-kiloparsec radius have been hypothesized to be an important mechanism for driving gas inflows to small scales, thus feeding central black holes. Recent numerical simulations have shown that the growth of central black holes in galaxies can destroy short bars, when the black hole reaches a mass of $\sim 0.1\%$ of the total stellar mass of the galaxy. We study $N$-body simulations of galaxies with single and double bars to track the long-term evolution of the central stellar mass distribution. We find that the destruction of the short bar contributes significantly to the growth of the bulge. The final bulge mass is roughly equal to the sum of the masses of the initial pseudo bulge and short bar. The initially boxy/peanut-shaped bulge of S\'ersic index $n\lesssim1$ is transformed into a more massive, compact structure that bears many similarities to a classical bulge, in terms of its morphology ($n \approx 2$), kinematics (dispersion-dominated, isotropic), and location on standard scaling relations (Kormendy relation, mass-size relation, and correlations between black hole mass and bulge stellar mass and velocity dispersion). Our proposed channel for forming classical bulges relies solely on the destruction of short bars without any reliance on mergers. We suggest that some of the less massive, less compact classical bulges were formed in this manner.
\end{abstract}


\section{Introduction} \label{sec:intro}

Galaxies are generally considered as composite stellar systems comprising a fast-rotating disk and a more slowly rotating bulge. How bulges form is an important topic in understanding the evolution of galaxies. Bulges come in two flavors. From a purely morphological perspective, classical bulges are highly concentrated, featureless spheroids, while pseudo bulges, characteristically hosted by late-type galaxies, are a more flattened, lower surface density component that often coexists with complex central substructures such as nuclear bars, disks, rings, and spirals \citep{kormendy2004secular}. Pseudo and classical bulges are thought to have completely different formation mechanisms. It is plausible that classical bulges are the end products of major mergers of gas-rich galaxies. Tidal torques drive efficient gas inflows during gas-rich mergers, which lead to rapid central (kpc-scale) starbursts and bulge build-up \citep[e.g.,][]{Hernquist1989Tidal, Barnes&Hernquist1996, Hopkins2009a}. Other processes, besides mergers, may also contribute to the growth of classical bulges. Zoom-in cosmological simulations suggest that misaligned accretion of gas can form bulges by \insitu\ starbursts, placing less emphasis on the role of mergers \citep{Scannapieco2009formation, Sales2012origin, Zolotov2015}. Gas-rich disks at high redshifts are gravitationally unstable to the formation of massive clumps, mergers of which provide yet another avenue to concentrate stars in the central regions of galaxies \citep{noguchi1998clumpy, Vandenberg1996age, noguchi1999early, Elmegreen&Elmegreen2005stellar, Genzel2008ring, Bournaud2008high, Dekel2009formation, Clarke&Debattista2019}. \citet{Elmegreen2008bulge} suggest that the clumps coalesce into a ``classical'' bulge. By contrast, \citet{Inoue&Saitoh2012nature} and \citet{du2015forming} argue that such clumps manufacture pseudo bulges/bars with significant rotation, bar-like morphology, and an exponential surface density profile. Even massive classical bulges can have diverse merger histories \citep{Bell2017galaxies}. \citet{Park2019horizon} suggest that roughly half of the spheroidal component in disk-dominated galaxies arises from orbits aligned with the disk; such disk stars continuously migrate to the center, without the aid of perturbations from mergers. \citet{Wang2019angular} similarly emphasized the contribution of disk star migration to bulge growth.


Pseudo bulges are thought to form through secular evolution of the large-scale disks of spiral galaxies, with bars being a key agent of angular momentum redistribution. Bars can funnel gas efficiently into the central regions of galaxies by bar-driven shocks, forming nuclear rings \citep[e.g.,][]{Athanassoula1992morphology, Kim2012gaseous, LiZhi2015hydrodynamical}, disks, or bars \citep{Shlosman1989Bars}. Nuclear disks are likely be classified as pseudo bulges morphologically. On the other hand, although boxy/peanut bulges are also classified as pseudo bulges, while they are likely be a part of bars that buckle vertically \citep[e.g.,][]{raha1991dynamical}. Additionally, some cosmological simulations \citep[e.g.,][]{Okamoto2013origin} suggest that pseudo bulges can form from the rapid supply of low angular momentum gas at $z\gtrsim2$, before the assembly of disks. In this case, pseudo bulges can be older than those forming in the secular evolution of disks ($z\lesssim 1$).

The empirical correlations between central black hole (BH) mass and bulge properties \citep{Ferrarese&Merritt2000, gebhardt2000relationship,Tremaine2002slope,Haring&Rix2004black, gultekin2009m} have prompted numerous suggestions that the two coevolve in some manner. However, as reviewed by \citet{kormendy2013coevolution}, the correlations are tight only for classical bulges and elliptical galaxies. Pseudo bulges exhibit a markedly larger scatter and lower zero point, and it is unclear what effect, if any, BHs have on their evolution. Indeed, the least massive central BHs known, with mass $M_{\rm BH} \approx 10^4 - 10^6\,M_\odot$, live in essentially bulgeless galaxies (\citealp{Filippenko&Ho2003lowmass, Barth2004POX52, Greene2008black, Jiang2011host}; see \citealp{Greene2019review} for a review, in preparation). BHs evidently do not require bulges to form \citep{Ho2008nuclear}. Further, the mere existence of disk-dominated active galaxies \cite[e.g.,][]{Cisternas2011bulk,Kim2017sellar,ZhaoDongyao2019role} implies that BHs can grow by internal secular processes alone.

Stellar bars can drive gas to sub-kpc scale efficiently, but transporting gas to yet smaller radii becomes challenging without the aid of smaller scale non-axisymmetric structures \citep{Hopkins&Quataert2010massive}, such as a short inner bar in double-barred (S2B) galaxies \citep{Shlosman1989Bars}. Roughly $1/3$ of barred galaxies in the local Universe are observed to be S2B galaxies \citep{erwin2002double, laine2002nested, erwin2004double}. \citet{Debattista2007} for the first time successfully generated a long-lived S2B structure in $N$-body simulations. The systematic study of \citet{du2015forming} demonstrated that a short (inner) bar can form spontaneously without involving gas from an initially dynamically cool, nuclear stellar disk due to its own bar instability \citep[see also][]{Wu2016time-dependent}. The addition of gas was considered by \citet{Wozniak2015double}. 

The inner short bar promotes the accretion onto the central BH, but the BH, in turn, mediates its own growth by destroying the bar \citep{Du2017black}. The destruction of bars under the dynamical influence of central massive concentrations (e.g., BHs) has long been studied \citep[e.g.,][]{gerhard1985triaxial,pfenniger1990dissipation,Hasan1990chaotic}. It is well known that an unrealistically massive BH ($>4\%$ of total stellar mass $M_\star$) is needed to destroy a large-scale bar \citep{Shen2004destruction, Athanassoula2005bars, Debattista2006secular}. \citet{Hozumi2012destructible} suggested that a weaker bar is not as robust as a large-scale bar. \citet{Du2017black} found a BH of mass $M_{\rm BH}\approx 10^{-3} M_\star$ destroys a short bar of 1 kpc scale quickly and hence suppresses its own growth. Thus, the maximum mass of BHs allowed in the secular evolution is about $10^{-3} M_\star$, which is consistent with observations \citep[e.g.,][]{kormendy2013coevolution,Reines2015relations}. In this paper, we define ``short bars'' as bars of radius $\lesssim 1.5$ kpc scale, no matter whether they co-exist with an outer bar. Thus, short bars can be the inner bar of S2B galaxies or small-size bar of single-barred galaxies. 

What is the remnant of the destroyed short bar? We expect the dissolved bar to become a denser, more axisymmetric structure. Can it be identified observationally? How does it affect the properties of bulges? Following on the work of \citet{Du2017black}, we here investigate in detail the properties of the remnant short bar. \refsec{sec:sim} describes our simulations. The process of morphological decomposition is presented in \refsec{sec:decomp}. The results of the decomposition and the intrinsic properties of the bulges are shown in \refsec{sec:result}. A physical scenario for the secular coevolution of BHs and bulges is discussed in \refsec{sec:picture}. We summarize our conclusions in \refsec{sec:sum}.

\section{Simulations}\label{sec:sim}

\subsection{Setup}

We use the method firstly presented by \citet{du2015forming} to generate galaxies with short bars: dynamically cold, rotation-dominated inner/nuclear disks are introduced in the central regions of pure-disk models. The simulations are run with the three-dimensional cylindrical polar grid option of the {\tt GALAXY} $N$-body code \citep{sellwood2014galaxy}, which increases the force resolution toward the center. The units system of the simulations is $G = M_0 = h_R = T_0 = V_0 = 1$, where $G, M_0, h_R, T_0,$ and $V_0$ are the units of the gravitational constant, mass, length, time, and velocity, respectively. We scale the models to mimic typical spiral galaxies by setting $M_0 = 4\times 10^{10} \, M_\odot$ and the initial scale length of the purely exponential disk to $h_R = 2.5$ kpc, which gives $T_0 = \sqrt{h^3_R/GM_0} \approx 9.3$ Myr and $V_0=\sqrt{GM_0/h_R}\approx 262\, {\rm km~s^{-1}}$. This scaling is the same as that used in \citet{du2015forming}. The simulation box measures $N_R \times N_\Phi \times N_z = 58 \times 64 \times 375$, which gives a force resolution of $\sim 25$ pc in the central regions. Such a grid can sufficiently resolve the dynamics of 1 kpc-scale short bars \citep{du2015forming}.
%

\begin{table}[htb]
\caption{\label{tab:param}%
Basic properties of the models.}
\begin{ruledtabular}
\begin{tabular} {lllll}
    Model\footnote{Name of the models. S2B\_a0, SB\_a0, and SB\_b0 are the control models of S2B\_a, SB\_a, and SB\_b, respectively. The only difference between each pair is the maximum mass of BHs $M_{\rm BH, max}$. In the control models, the black holes have little effect on the bar evolution.} &$b_{Q_{\rm min}}$\footnote{The minimum value of Toomre $Q$ at the center, quantifying the dynamical temperature of initial inner disk.} & $M_{\rm BH, max}\, (M_\star)$  
    &$R_{\rm bar}\,({\rm kpc})$\footnote{Half-size radius of inner/short-scale (left column) and outer/large-scale (right column) bar measured by the minimum radius obtained from tracing half-way down the peak of the $m=2$ amplitude (\reffig{fig:bar_fourier}) and the $10^{\circ}$ deviation from a constant phase angle at $t=2.8$ Gyr in models S2B\_a(0), SB\_a(0), and SB\_b(0), when the BH has no significant effect.}
    &  Type\footnote{S2B: double-barred galaxy; SB: single-barred galaxy. Note that in SB\_a a weak, longer bar forms after the BH destroys the short bar.} \\
    \colrule
    S2B\_a & 0.5 & $2\times 10^{-3}$ & $0.9$ \ \ $ 7.0$ & S2B \\
    SB\_a & 0.7  & $2\times 10^{-3}$ & $ 1.5$ \ \  ---  & SB  \\
    SB\_b & 0.9  & $2\times 10^{-3}$ & --- \ \ \  $ 6.3$         & SB  \\
    \colrule
    S2B\_a0 & 0.5 & $10^{-4}$ & $0.9$ \ \ $7.0$ & S2B \\
    SB\_a0 & 0.7  & $10^{-4}$ & $1.5$ \ \ ---  & SB  \\
    SB\_b0 & 0.9  & $10^{-4}$ & --- \ \ \  $6.3$         & SB  \\
\end{tabular}
\end{ruledtabular}
\end{table}

\begin{figure*}[htb]
    \centering
    \includegraphics[width=\linewidth]{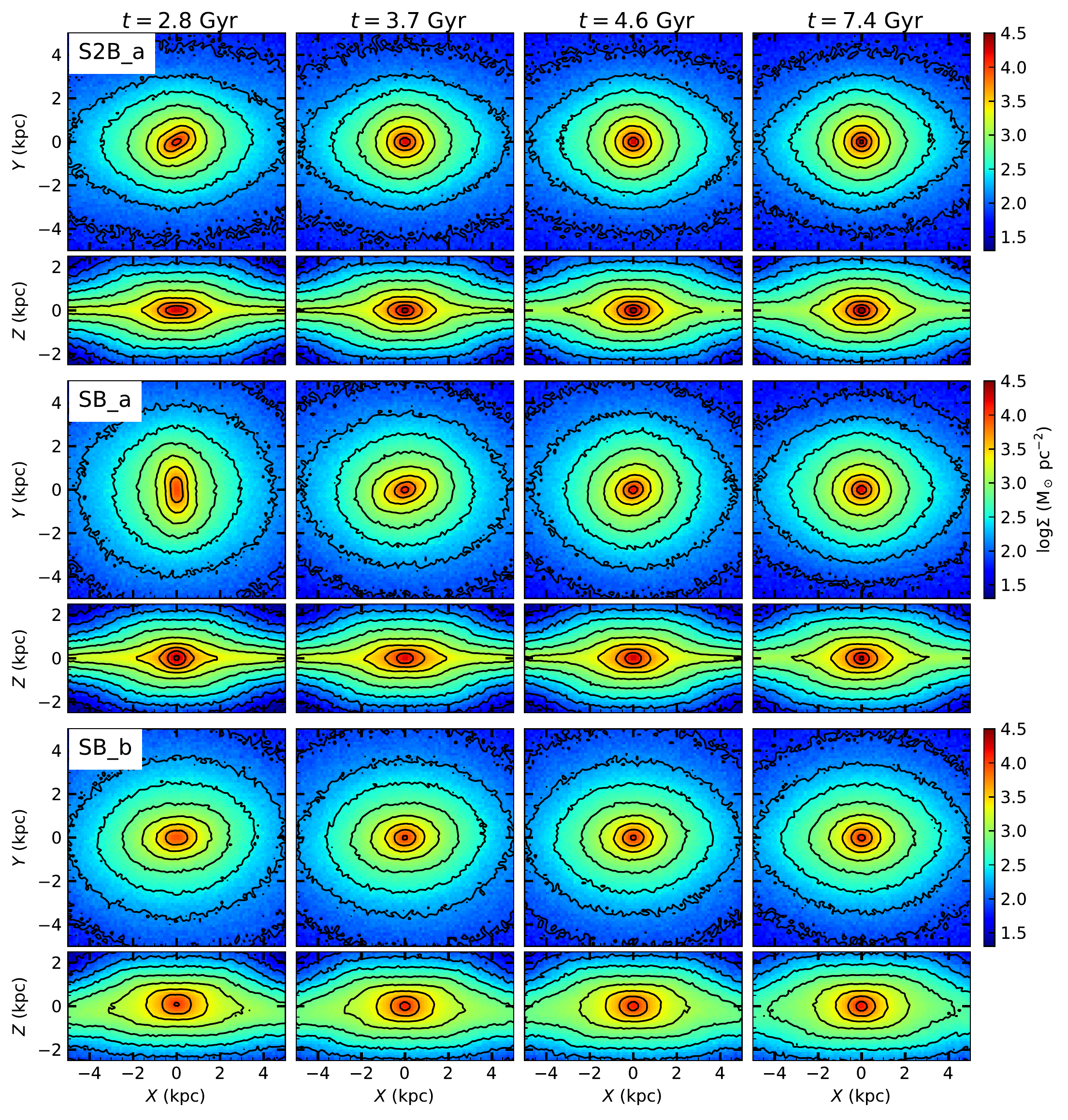}
    \caption {Face-on (upper panels) and edge-on (lower panels) surface density distributions of models S2B\_a, SB\_a, and SB\_b (from top to bottom). From left to right, we show the cases at time $t=2.8, 3.7, 4.6$, and 7.4 Gyr, respectively. All the images are presented using the same color bar and contours. The outer bars are aligned with the $x$-axis. These images clearly show that the short bars of models S2B\_a and SB\_a are destroyed by the BH, forming a spheroidal, denser structure in the central region.}
    \label{fig:model}
\end{figure*}

The basic properties of the models are given in \reftab{tab:param}. All of the models are initially composed of a live disk, a rigid dark matter halo, and a tiny BH. To simplify the simulations, we use rigid potentials to mimic dark matter halos, as the central dynamics are largely dominated by the stellar component. The halo potential is logarithmic, $\Phi(r) = 0.5 V_h^2\ln{(r^2+ r_h^2)}$, where $V_h=0.6V_0$ and $r_h=15h_R$. The purely exponential disk of mass $M_\star = 1.5M_0$ and initial scale length $h_R$ consists of four million equal-mass particles. Their gravitational force is softened with a radius of $0.01h_R=25$ pc. We use an additional potential of Plummer form, $\Phi_{\rm BH}(r)=-GM_{\rm BH}(t)/\sqrt{r^2+\epsilon_{\rm BH}^2}$, where $\epsilon_{\rm BH}=0.01h_R=2.5$ pc, to represent the central BH. The BH mass, $M_{\rm BH}$, is the same as the stellar particles before $t=300T_0=2.8$ Gyr, thus having no effect on the overall evolution of the models. Then it grows smoothly and adiabatically over 50 time units ($\sim0.5$ Gyr) from the level of stellar particles to a maximum value $M_{\rm BH, max}$, following a cosine function \citep[see details in][]{Du2017black}. 
%
%
%
$M_{\rm BH, max}$ used in each model is given in the third column of \reftab{tab:param}. At later times, $M_{\rm BH}$ is kept constant at $M_{\rm BH, max}$. It is worth mentioning that the force from the BH is added to each particle from the analytic form, and is therefore independent of the grid resolution. The guard shell technique is employed to reduce the time steps of gravitational integration around the BH \citep[see details in][]{Shen2004destruction, Du2017black}. 

\begin{figure*}[hbt]
    \centering
    \includegraphics[width=\linewidth]{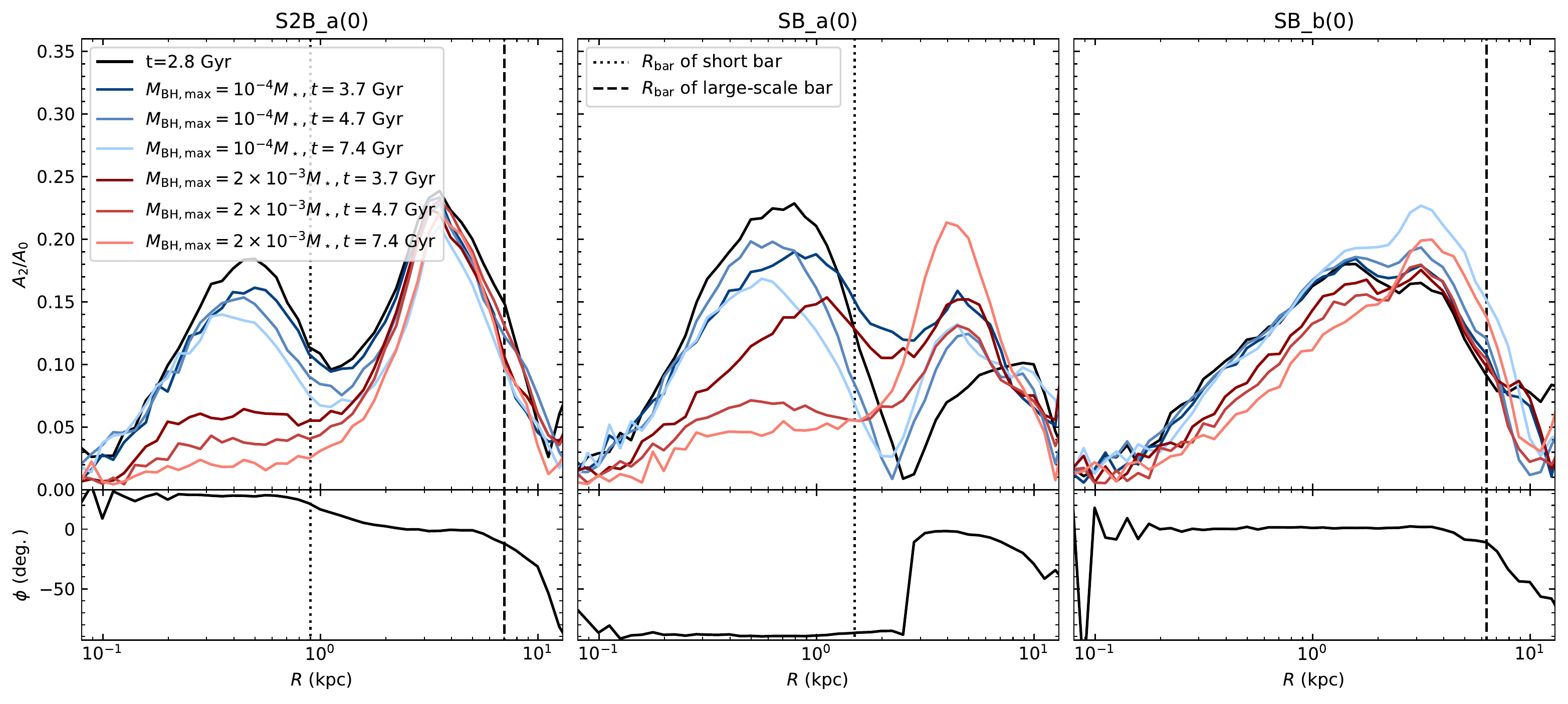}
    \caption{Time evolution of $A_2/A_0$ in models S2B\_a(0), SB\_a(0) and SB\_b(0) (from left to right), where $A_2$ and $A_0$ are the Fourier $m=2$ and $m=0$ modes, respectively, measured in annuli of equal radial interval in logarithmic space. The red and blue profiles correspond to the models of BH mass $M_{\rm BH,max}=2\times 10^{-3}\,M_\star$ and $10^{-4}\,M_\star$, respectively, at $t\geq 3.7$ Gyr. The black profile in each panel is the case of $t=2.8$ Gyr. The bottom panels show the phase angle $\phi$ of the bars at $t=2.8$ Gyr. The sizes of short/inner and large-scale/outer bars are marked by the dotted and dashed vertical lines, respectively.}
    \label{fig:bar_fourier}
\end{figure*}

The models are named S2B\_a(0), SB\_a(0), and SB\_b(0), according to their bar structures (see \reffig{fig:model} for the face-on and edge-on surface density distributions). The nomenclature S2B\_a(0) represents both S2B\_a and S2B\_a0, and likewise for SB\_a(0) and SB\_b(0). S2B\_a(0) has a double-barred structure. \citet{du2016kinematic} studied the kinematic properties of model S2B\_a0 and concluded that they are consistent with observed S2Bs. Thus, S2B\_a(0) was used as the standard model of S2Bs in \citet{du2015forming, du2016kinematic, Du2017black}. In this paper, we include new models SB\_a(0) and SB\_b(0) that have exactly the same halo and BH as S2B\_a(0). Their main difference is in their bars. Both SB\_a(0) and SB\_b(0) have only a single bar, while the bar in SB\_a(0) of radius $\sim 1.5$ kpc is much shorter than that of SB\_b(0); it is, in other words, a short bar (see bar size, $R_{\rm bar}$, in \reftab{tab:param}). We approximate the bar size---marked by vertical lines in \reffig{fig:bar_fourier}---as the minimum radius obtained by tracing half of the amplitude of the peak of the $m=2$ Fourier component ($A_2/A_0$) and the $10^{\circ}$ deviation from a constant $\phi$ at $t=2.8$ Gyr. Note that, in SB\_a(0) at $t=2.8$ Gyr, the weak $m=2$ component having $A_2/A_0\approx 0.1$ and ellipticity $\epsilon \approx 0.15$ at $r\approx7 \ {\rm kpc}$ is not strong enough to be considered as an outer bar.

The formation of bars is largely determined by Toomre's (1964) $Q$ parameter of the disk. We set $Q\approx2$ in the outer part of the disk; in the inner part ($R\lesssim 4.4$ kpc), $Q$ is reduced gradually toward the center. This results in a dynamically cool inner disk, reaching a minimum value $b_Q$ at the center. Thus, we use $b_Q$ (\reftab{tab:param}) to represent the dynamical temperature of the inner disk. The dynamically cool inner disk leads to the formation of the inner bar in model S2B\_a(0) and the short bar in model SB\_a(0) \citep[see more in][]{du2015forming}. In the following section, we present the evolution of these models under the dynamical influence of the BH. This work only considers the dynamical effect of the BH; no hydrodynamic processes are included.

%
%
%

\begin{figure*}[htb]
    \centering
    \includegraphics[width=0.9\linewidth]{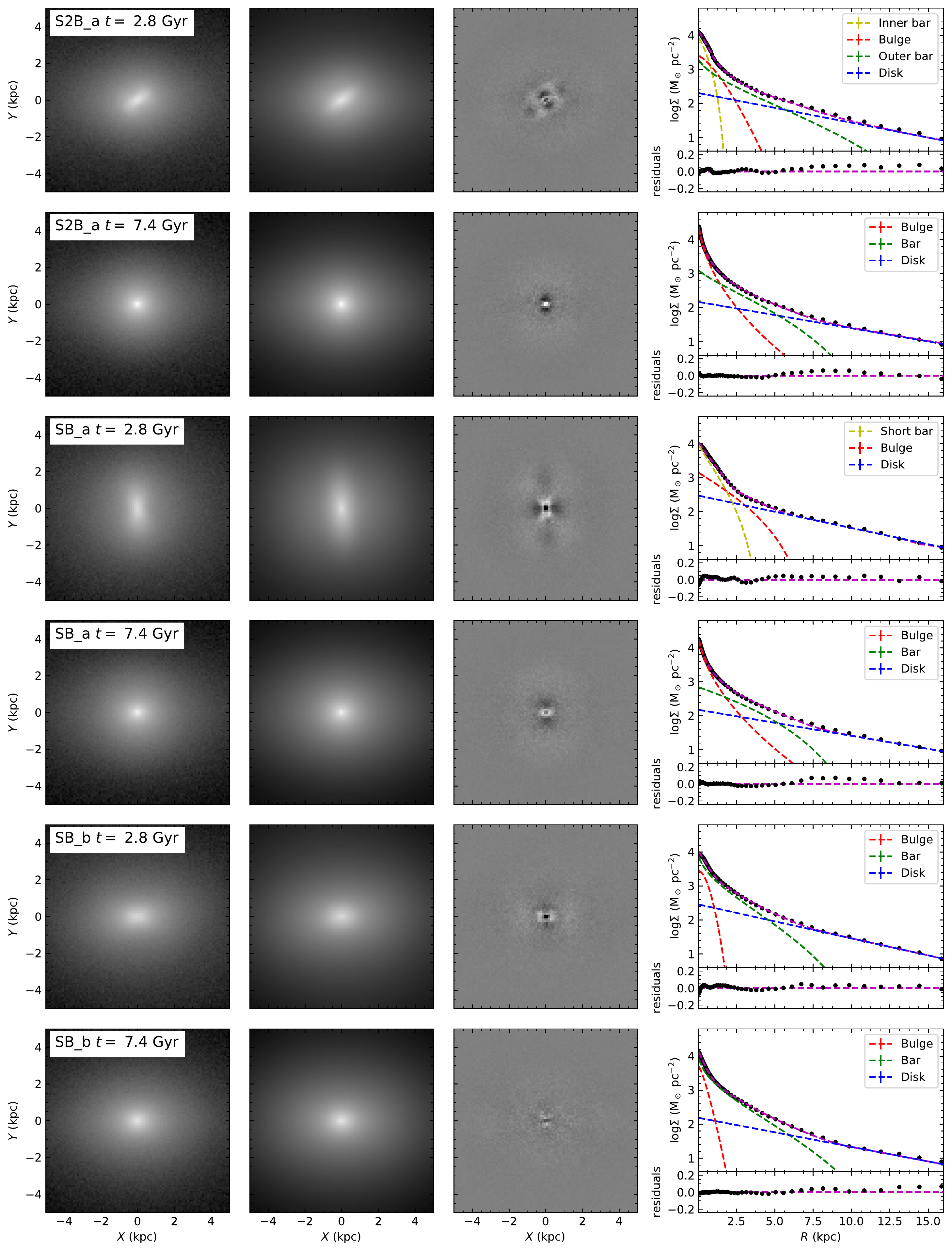}
    \caption{Morphological decomposition using \galfit. From top to bottom, we show the three models at $t=2.8\ \rm Gyr$ and $t=7.4\  \rm Gyr$, which correspond, respectively, to the time when the BH starts to growth and the final time of each simulation. From left to right, we show images of the models, the \galfit\ fitting, the residuals, and the 1D profiles, respectively. The field of view is $10\times10$ kpc$^2$. The images are shown using the same logarithmic stretch for the model and fitting image, and histogram equalization stretch is used for the residual image. In the right panels, we show the 1D density profiles of each component used in the \galfit\ model. The magenta lines correspond to the overall density profiles of the \galfit\ fit. The residuals are shown in the lower part of the right panels. Both the bulges and bars are described by S\'ersic profiles; modeling the bar with a Ferrers function gives similar results.}
    \label{fig:decom_all}
\end{figure*}

\subsection{Evolution: the destruction of short bars due to the growth of black holes}

Figures \ref{fig:model} and \ref{fig:bar_fourier} show the evolution of models S2B\_a(0), SB\_a(0), and SB\_b(0). At $t\leq 2.8$ Gyr, models S2B\_a0, SB\_a0, and SB\_b0 are identical to S2B\_a, SB\_a, and SB\_b, respectively. The BH masses of the main group (S2B\_a, SB\_a, SB\_b) increase to $M_{\rm BH, max}=2\times 10^{-3}\,M_\star$, reaching a typical observed BH mass fraction. For comparison, the BH is unimportant ($M_{\rm BH, max}=10^{-4}\,M_\star$) over the entire simulation of the control group (S2B\_a0, SB\_a0, SB\_b0). 

The evolution of the models can be separated into three phases: 
\begin{enumerate}
    \item[(1)] {\it Bars and boxy/peanut-shaped bulges form spontaneously.} At $t\leq 2.8$ Gyr, all of the models form bars and boxy/peanut bulges spontaneously due to their internal dynamical instabilities (the first column of Figure \ref{fig:model}). All bars have reached steady-state. At this stage the BH is still small and has no significant effect on the evolution of their hosts. As shown in the edge-on images in \reffig{fig:model}, boxy/peanut bulges form due to the buckling instability triggered by bars, probably large-scale bars.

    \item[(2)] {\it Short (inner) bars are destroyed due to the growth of the BH.} At $2.8<t \leq 3.3$ Gyr, the BH grows smoothly to the maximum mass $M_{\rm BH, max}$. The short bars of S2B\_a and SB\_a are completely destroyed by the BH, in about $0.4$ Gyr and $1.4$ Gyr (the second and third columns of \reffig{fig:model}), respectively. A spheroidal component forms in the central region, where $A_2/A_0$ decreases significantly (the series of red profiles in \reffig{fig:bar_fourier}). Using $M_{\rm BH,max}=10^{-4}\,M_\star$ produces a minor effect in models S2B\_a0 and SB\_a0, with the short bars surviving until the end of the simulations (the series of blue profiles). In SB\_b, the central region of the bar becomes rounder, but this effect is not as pronounced as in both S2B\_a and SB\_a. 

    \item[(3)] {\it Steady phase.} After the short bars are destroyed, the galaxies evolve slowly. The morphology is unchanged until the end of the simulation.
\end{enumerate}

In order to study the properties of the remnant of the short bar destruction, we decompose the models in \refsec{sec:decomp}. The changes in morphology are investigated from an observational point of view.

\begin{figure*}[htb]
    \centering
    \includegraphics[width=\linewidth]{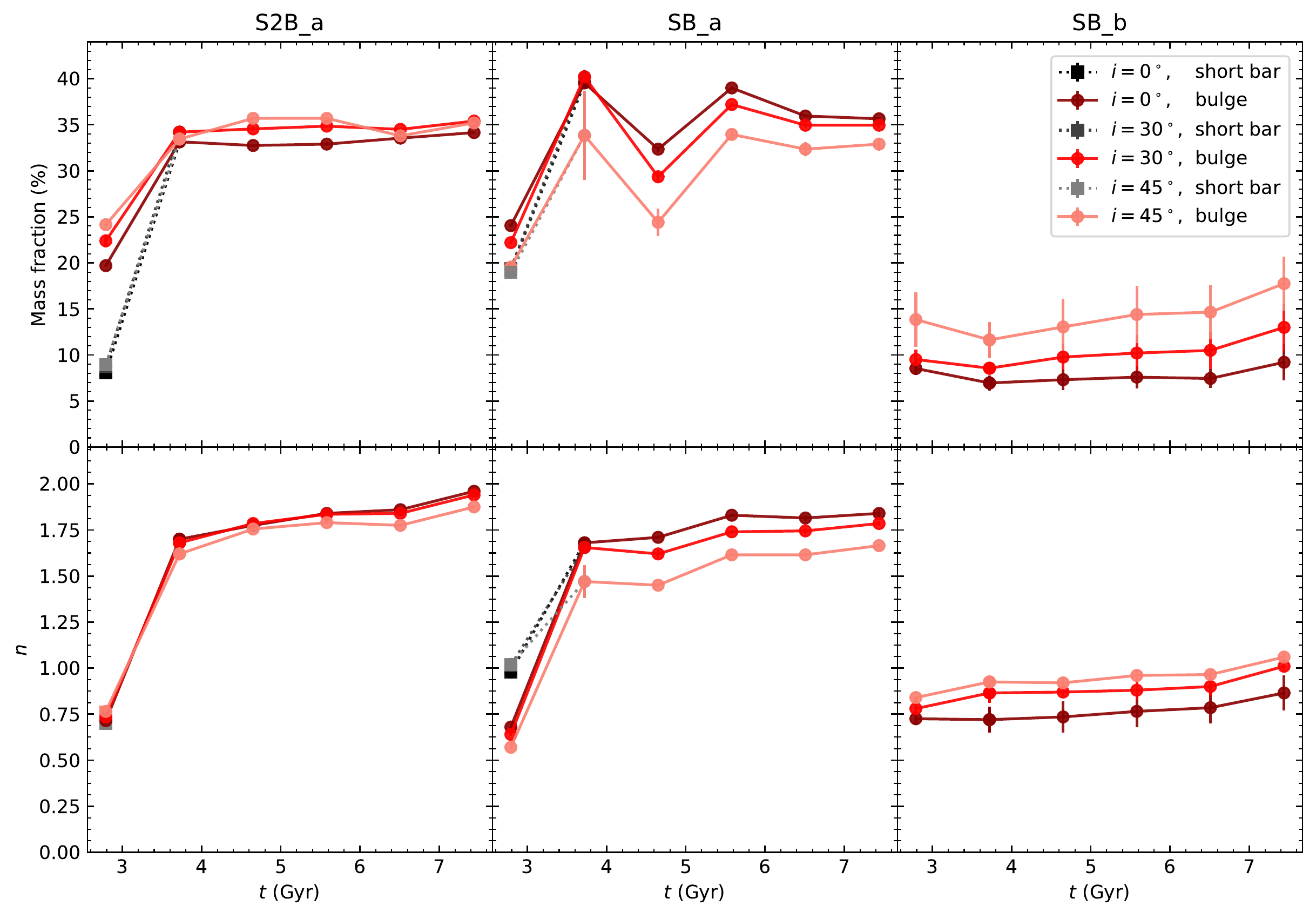}
    \caption{Evolution of the mass fraction (top) and S\'ersic index $n$ (bottom) of the bulges in S2B\_a, SB\_a, and SB\_b (from left to right). The results of the short bars are overlaid for models S2B\_a and SB\_a. 
The error bars represent the results obtained using Ferrers and S\'ersic bars; the dots mark their average values. The error bars are generally smaller than the size of the symbols, except for the case of $i=45^\circ$. The squares show the mass fraction and S\'ersic index of the short bars. From dark to light colors, we vary the inclination angle from $i = 0^\circ$ to $45^\circ$. }
    \label{fig:evol_all}
\end{figure*}

\begin{figure*}[htb]
    \centering
    \includegraphics[width=0.9\linewidth]{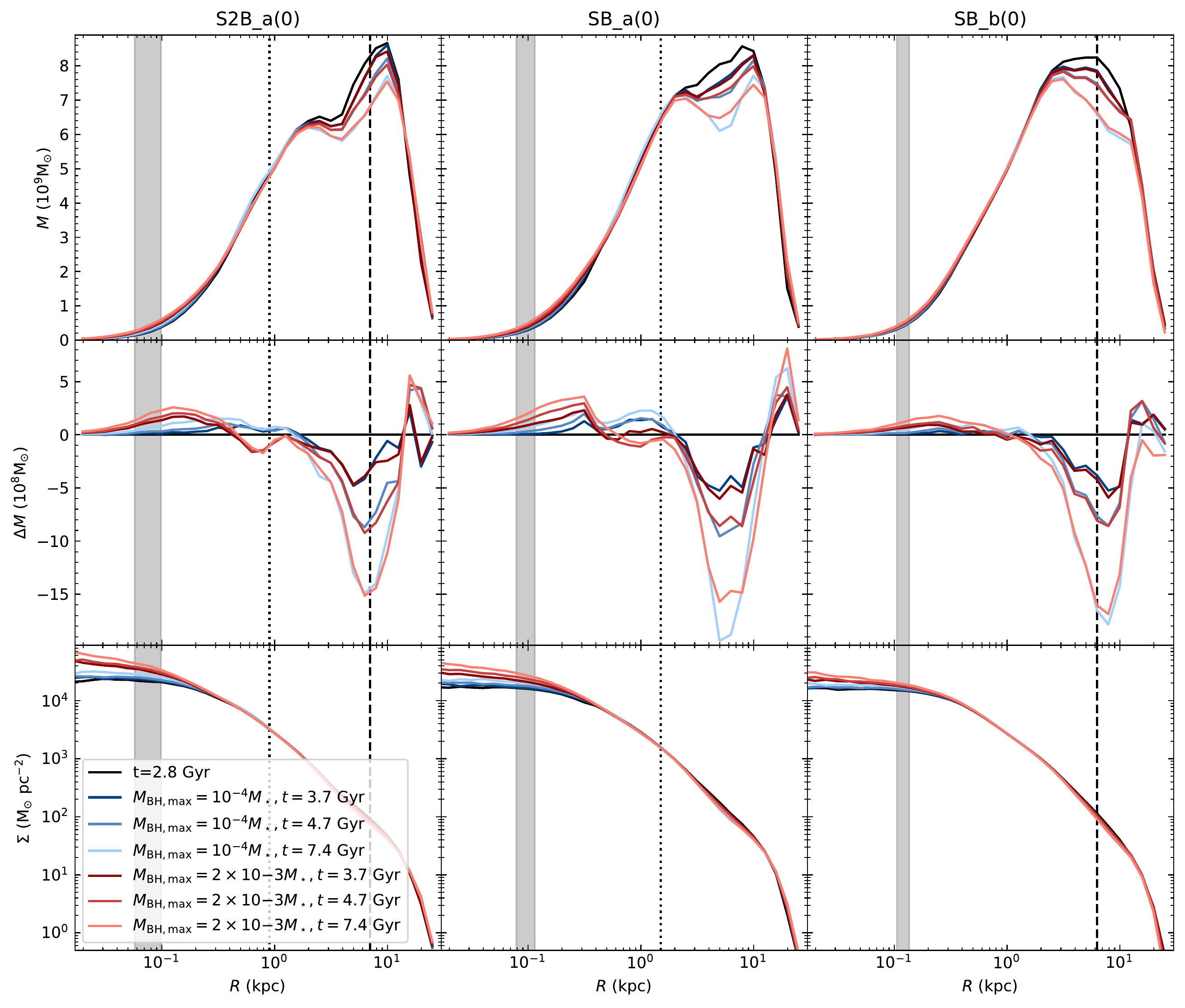}
    \caption{Top: mass distribution per logarithmic bin of radius $R$. Middle: the difference of mass distribution between $t=2.8$ Gyr and the other time steps. Bottom: surface density profiles. From left to right, the three models [S2B\_a(0), SB\_a(0), and SB\_b(0)] are shown. The red and blue series of profiles represent models with BHs of $M_{\rm BH,max}=2\times 10^{-3}\,M_\star$ and $M_{\rm BH,max}=10^{-4}\,M_\star$, respectively. The grey region shows the range of the sphere-of-influence of the BH, within which the stellar mass is equal to the BH mass. The sizes of short/inner bars and large-scale/outer bars are marked by dotted and dashed vertical lines.
    }
    \label{fig:line_mL}
\end{figure*}

\begin{figure*}[htb]
    \centering
    \includegraphics[width=0.9\linewidth]{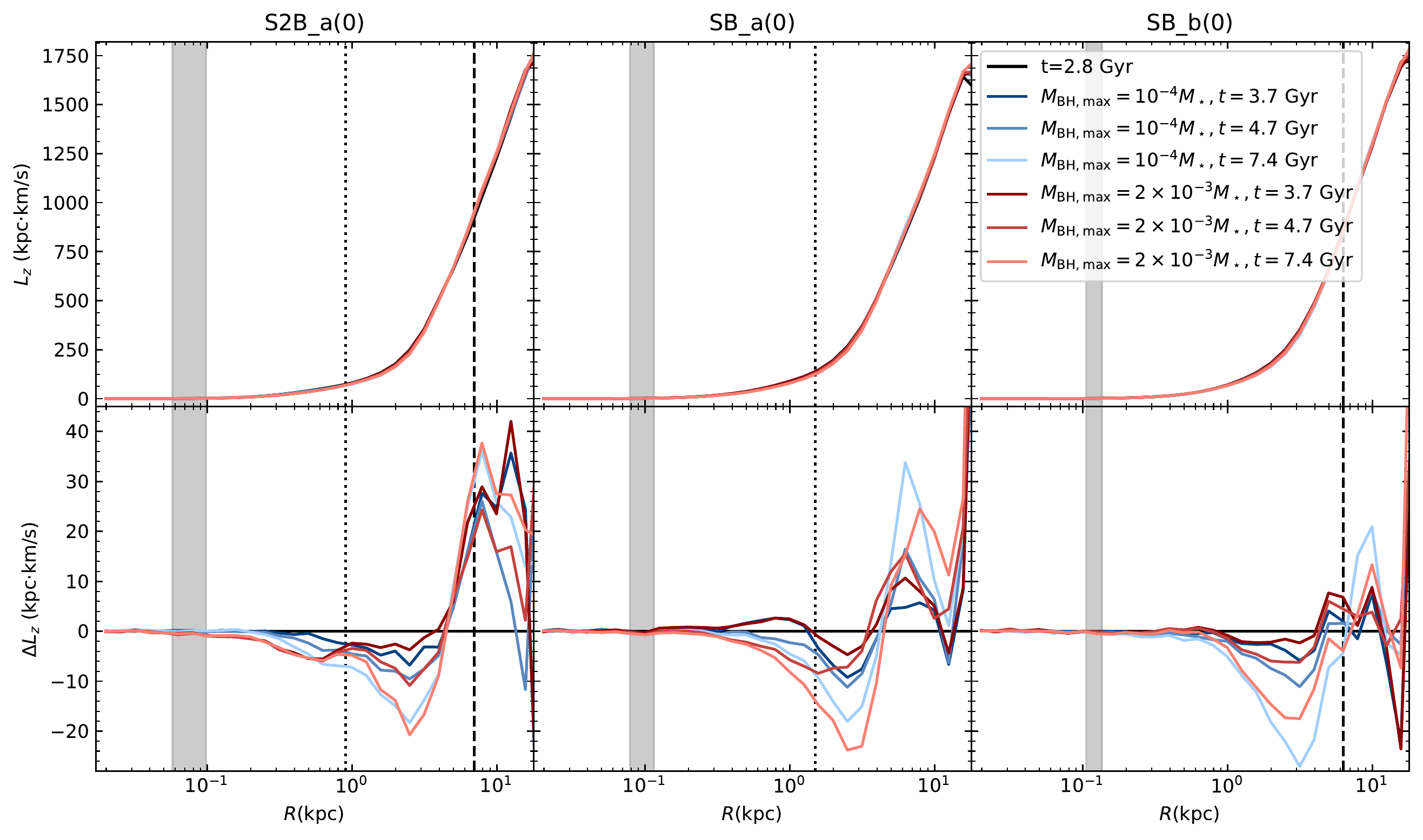}
    \caption{Similar to \reffig{fig:line_mL}, the distribution of specific angular momentum ($L_z$; top) per logarithmic bin of radius and the difference between $t=2.8$ Gyr and the other time steps ($\Delta L_z$; bottom). The red and blue series of profiles represent the models with BHs of $M_{\rm BH,max}=2\times 10^{-3}\,M_\star$ and $M_{\rm BH,max}=10^{-4}\,M_\star$, respectively.
    }
    \label{fig:line_L}
\end{figure*}

\section{Morphological Decomposition}\label{sec:decomp}
\subsection{Setup for {\tt GALFIT}}

We employ the latest version of \galfit\ \citep{peng2002detailed, peng2010detailed}---a widely used standard tool for decomposing galaxy images---to investigate the morphological structures of the models. We generate mock observational images using a Cartesian grid covering a region of $30 \times 30$ kpc$^2$. Each cell has an equal size of $ 0.1 \times 0.1$ kpc$^2$ that is sufficient for decomposing all structures. To test the effect of inclination, we project all of the models to typical inclination angles of $i=0^\circ$ (i.e. face-on, \reffig{fig:decom_all}), $30^\circ$, and $45^\circ$. We do not account for the effects of sky background or point-spread function.

All of the models are centered at the coordinate origin. The azimuthal shape of each component is the pure ellipse

\begin{equation} 
r(x,y) = \sqrt{x^{2}+\Big(\frac{y}{1-\epsilon}\Big)^{2}},
\end{equation}

\noindent
where $\epsilon$ is the ellipticity. We use a simple exponential profile to fit the disk component,

\begin{equation}
\Sigma (r)=\Sigma_0 \exp \Big{(}-\frac{r}{r_s}\Big{)},
\end{equation}

\noindent
where $r_s$ is the scale length and $\Sigma_0$ the central surface density. Bulges are described by the S\'ersic function 

\begin{equation}
\Sigma (r)=\Sigma_e \exp \Big{[}-\kappa\Big{(}\big{(}\frac{r}{r_e}\big{)}^{1/n}-1\Big{)} \Big{]},
\end{equation}

\noindent
where $r_e$ is the half-mass (effective) radius, and $\Sigma_e$ is the surface density at $r_e$. The S\'ersic index $n$ is generally used to represent the concentration, and $\kappa$ satisfies $\Gamma(2n)=2\gamma(2n,\kappa)$, where $\Gamma$ and $\gamma$ are the gamma function and incomplete gamma function, respectively.

The bar is fit using both the S\'ersic function and the modified Ferrers function, which is give as 

\begin{equation}
\Sigma (r)=\Sigma_0 \Big{(}1-(r/r_{\rm out})^{2-\beta}\Big{)}^\alpha,
\end{equation}

\noindent
where $r_{\rm out}$ is the radius of the outer truncation, $\Sigma_0$ is the central surface density, and $\alpha$ and $\beta$ control the sharpness of the outer truncation and the central concentration, respectively.

\subsection{Decomposition of individual models}

\reffig{fig:decom_all} shows the morphological decomposition of the face-on images of models S2B\_a, SB\_a, and SB\_b. The columns show, respectively, the logarithmic surface densities of the models, the fitting results obtained by \galfit, and the residuals. The BH is tiny at $t=2.8$ Gyr, the starting point of Phase 2. The snapshots at $t=7.4\ \rm Gyr$ represent the morphological decomposition after the short bars have been destroyed. During the steady phase (Phase 3), the morphologies of all models show little variation. It is clear from the one-dimensional (1D) radial density profiles of the individual components (fourth column) that all models are well fitted (residuals $<0.05$). Each model includes an exponential disk, a bulge, and one or two bar components. The bulge components possibly correspond to the boxy/peanut-shaped bulges that are clearly seen in the edge-on plots of \reffig{fig:model}. Two bars are required in order to fit S2B\_a(0) at $t=2.8$ Gyr (\reffig{fig:decom_all}); after the short bar is destroyed, only one bar is used (e.g., the cases at $t=7.4$ Gyr). The single-barred models SB\_a(0) and SB\_b(0) always include one bar component. Note that in SB\_a a weak, longer bar forms after its short bar is destroyed. In comparison, SB\_a0 only has a single short bar, as the ellipticity of its outer disk is always $<0.2$. The peak at $R\simeq 4 $ kpc (small peak of \reffig{fig:bar_fourier}) corresponds to the newly formed outer bar in SB\_a. The value of $A_2/A_0$ increases from 0.15 to 0.21 during $3.7-7.4$ Gyr, perhaps due to the destruction of the short bar.

%
\begin{figure*}[htb]
    \centering
    \includegraphics[width=0.9\linewidth]{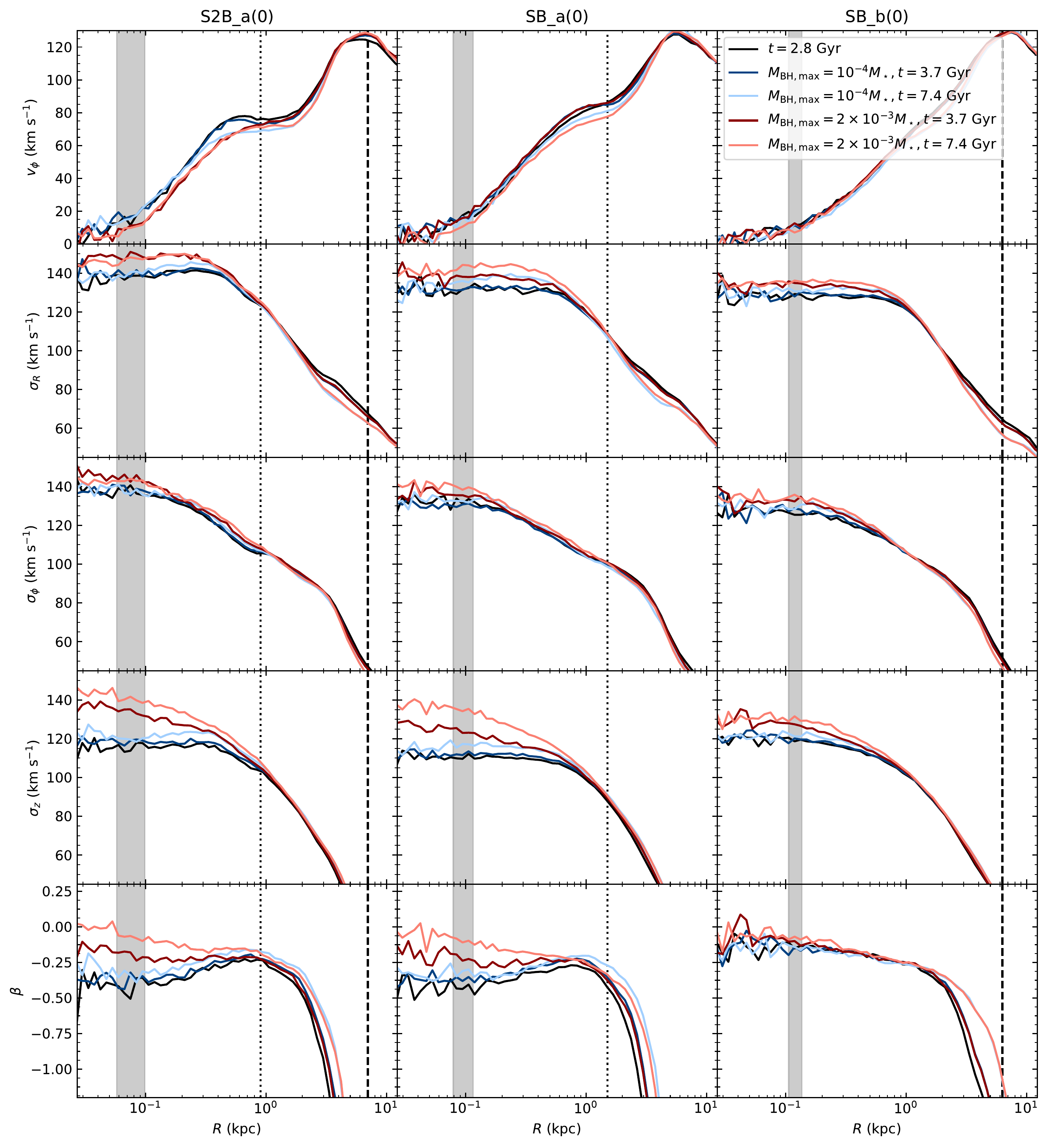}
    \caption{Similar to Figure 5, evolution of rotational velocity ($v_\phi$), velocity dispersions ($\sigma_R,\sigma_{\phi}$, $\sigma_z$), and anisotropy parameter [$\beta \equiv 1-(\sigma^2_\phi+\sigma^2_R)/2\sigma^2_z$], measured in annuli of equal radial interval in logarithmic space. The red and blue series of profiles represent the models with BHs of $M_{\rm BH,max}=2\times 10^{-3}\,M_\star$ and $M_{\rm BH,max}=10^{-4}\,M_\star$, respectively. The grey region in each panel shows the range of the sphere-of-influence of the BH. The sizes of short bars and outer bars are marked with dotted and dashed vertical lines. 
    }
    \label{fig:line_sigma}
\end{figure*}

\begin{figure}[htb]
    \centering
    \flushleft
    \includegraphics[width=\linewidth]{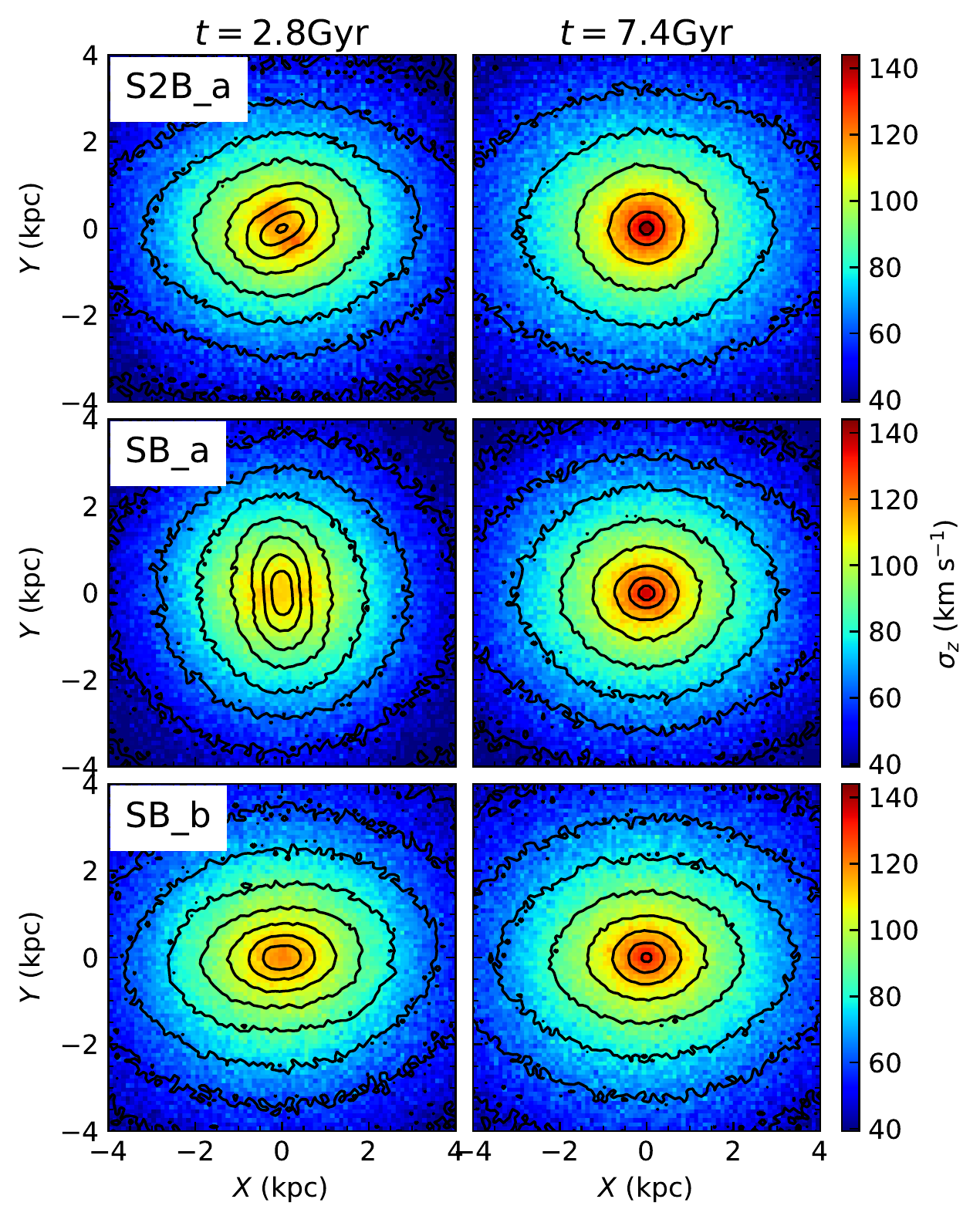}
    \caption{Evolution of the vertical velocity dispersion ($\sigma_z$), viewed face-on. The contours correspond to the distribution of surface density. At $t=2.8$ Gyr, $\sigma$-humps (hollows) 
     are perpendicular (parallel) to the short bars in both models S2B\_a and SB\_a.
    }
     \label{fig:sigma_map}
     %
\end{figure}

\section{Results} \label{sec:result}

This section shows the results of the morphological decomposition. We investigate the evolution of the bulges, not only in terms of morphology, but also the intrinsic exchange of mass and angular momentum.

\subsection{The growth of bulges after short bar destruction}
\label{sec:resultgalfit}

\reffig{fig:evol_all} illustrates the evolution of the mass fraction and S\'ersic index of the bulges in S2B\_a, SB\_a, and SB\_b. The effect of inclination angle is tested in the range $i = 0^\circ-45^\circ$. We use the upper and lower limits of the error bars to represent the results obtained by using Ferrers and S\'ersic bars, respectively; the dot symbols mark their average values. Ferrers function generally gives more massive bars than the S\'ersic function, as a result of which the bulge mass is slightly larger when a S\'ersic bar is used. For models S2B\_a and SB\_a, we overlay the properties of their short bars at $t=2.8$ Gyr (square symbols), after which the short bars are destroyed quickly. The dotted lines indicate a potential evolution track by which the remnants of short bars contribute to the growth of bulges. In comparison, the bulge properties of the control group evolve mildly during $t>2.8$ Gyr (not shown here).

The bulge masses of S2B\_a and SB\_a clearly increase significantly after their short bars are destroyed. The inner bar of S2B\_a ($R_{\rm bar}\approx 0.9$ kpc, mass $\sim0.08\,M_\star$) is destroyed in $\sim 0.4$ Gyr during Phase 2 (\refsec{sec:sim}), while for model SB\_a it takes $\sim 1.4$ Gyr to destroy the short bar, which is longer ($R_{\rm bar}\approx 1.5$ kpc) and more massive ($\sim 0.2\, M_\star$). Although the two short bars differ greatly in strength, they evolve similarly: a massive ($\sim 0.35\,M_\star$), centrally concentrated (S\'ersic index $n\approx 1.75$) bulge forms in the aftermath of the destruction of the short bar. As a consequence, the formerly boxy/peanut-shaped bulge (with $n<1$) takes on more of an appearance of a classical bulge. The S\'ersic index increased to a value close to $2$ without involving any external perturbation. A S\'ersic index of $n\gtrsim 2$ is generally used as a criterion for defining classical bulges (\citealp{Fisher&Drory2008AJ....136..773F}; but see \citealp{Gao2019demographics}). For comparison, the bulge of the single-barred case (SB\_b; right panels) has changed little. This is consistent with the numerical simulations of \citet{Debattista2004bulge} showing that box/peanut bulges are able to maintain a low S\'ersic index ($n<1.5$). All of the results above are roughly independent of inclination angle and choice of bar function. After $t=4$ Gyr (Phase 3), all the bulges evolve slowly. 

The particles of the destroyed short bars become incorporated as part of the bulges. The bulge masses of both S2B\_a and SB\_a are similar to the sum of the masses of their short bars and progenitor bulges. There are residual differences at the level of $\sim 0.1\,M_\star$, which may reflect inward mass transport or the uncertainty of the morphological decomposition. About half of the mass of the resultant bulges is from the progenitor bulges. The properties of resultant bulges are largely determined by both the progenitor bulges and the relics of the short bars.

\subsection{Redistribution of mass and angular momentum}

To understand the growth of the bulges, we investigate the transport of mass ($M(t)$; Figure \ref{fig:line_mL}) and angular momentum ($L_z(t)$; Figure \ref{fig:line_L}) during $t = 2.8-7.4$ Gyr. The profiles of $M(t)$ and $L_z(t)$ (first row) and their differentials between different epochs ($\Delta M$ and $\Delta L_z$; second row) are measured in annuli of equal radial interval in logarithmic space. \reffig{fig:line_mL} further shows the evolution of the radial surface density profile, which is pronounced within the BH's sphere-of-influence. 

The effect of the destruction of the short bar can be seen clearly by comparing the main group of models with $M_{\rm BH, max} = 2\times 10^{-3}\,M_\star$ (red profiles) with the control group with $M_{\rm BH, max}=10^{-4}\,M_\star$ (blue profiles). Consistent with \citet{Athanassoula2004bars} and \citet{Debattista2006secular}, stars are transported outward by gaining angular momentum around the corotation radius of the outer bar. Thus, the regions of $\Delta M<0$ and the turning point of $\Delta L_z$ are roughly consistent with the outer ends of the bar (marked by vertical dashed lines). This mechanism is efficient even in the case of the extremely weak bar at $R\approx 5$ kpc for SB\_a(0), and the behavior is very similar for both the main and the control groups. We confirmed that the outer bars are fast bars, based on the criterion of \citet{Debattista2002fast}. There is no clear signature of outward mass transfer around the ends of the short bars, possibly because the short bars, being slow, are much shorter than their corotation radii ($\sim 3.5$ kpc; \citealp{du2015forming}), rendering angular momentum exchange inefficient. In the control group, the mild increase of mass ($\Delta M>0$) in the central region might be partially due to long-term asymmetric drift or the inward migration of stars that lose angular momentum around the outer bar's corotation radius. Models S2B\_a and SB\_a ($M_{\rm BH, max}=2\times 10^{-3}\,M_\star$) apparently transfers more mass from $\sim 1$ kpc to $<300$ pc due to the destruction of the short bar during $2.8-3.7$ Gyr (Phase 2). As a consequence, the central surface density (the third row of \reffig{fig:line_mL}) becomes cuspier. However, the additional mass transport is only about $1\%-2\%$ of $M_\star$. The red series of $\Delta L_z$ profiles have smaller values at $R<1$ kpc, suggesting that angular momentum is transferred outward as a result of short bar destruction, although this effect is not significant. After $t=3.7$ Gyr, the changes in mass and angular momentum are minor at the central regions. 

To summarize: the destruction of the short bar contributes to the growth of bulges. In this scenario, a massive and compact bulge forms by absorbing the stars of the short bar destroyed by BHs. A nuclear cusp forms, leading to a larger S\'ersic index, although only $\sim 1\%-2\%$ of the total stellar mass is transferred inward from $\sim 1$ kpc. \citet{Du2017black} argued that the growth of the BH can be mediated by the secular evolution of short bars. The present study shows that, in return, the BH regulates the growth of the bulge, the two acting as a self-regulated system.

\subsection{Kinematics}

The radial profiles of the cylindrical rotation velocity ($v_\phi$) and the velocity dispersions ($\sigma_R, \sigma_{\phi}$, $\sigma_z$) indicate that the models are dominated by random motion in their central region (\reffig{fig:line_sigma}). The destruction of the short bar causes $\sigma_z$ to increase sharply in the center of S2B\_a and SB\_a, while $\sigma_\phi$ and $v_\phi$ are only mildly affected. This may reflect the random scattering of bar orbits by the BH. The two-dimensional maps of $\sigma_z$ (\reffig{fig:sigma_map}) show that $\sigma_z$-humps/hollows\footnote{For details of $\sigma$-humps/hollows, see \citet{deLorenzo-Caceres2008} (integral-field unit observations), \citet{du2016kinematic} ($N$-body simulations), and \citet{du2017sigma} (theoretical interpretation).} \citep{deLorenzo-Caceres2008, du2016kinematic} disappear after the short bar is destroyed. The central peak in $\sigma_z$ occurs at $t=7.4$ Gyr, at which point the central region is also nearly isotropic, as judged by the anisotropic parameter, $\beta\equiv 1-(\sigma^2_\phi+\sigma^2_R)/2\sigma^2_z \approx 0$ (\reffig{fig:line_sigma}). Thus, a spheroidal structure dominated by random motions---a bulge---is created after the short bar is destroyed.

\subsection{Evolution of the bulges on scaling relations}\label{sec:scaling_relation}

\begin{figure*}[htb]
    \centering
    \includegraphics[width=\linewidth]{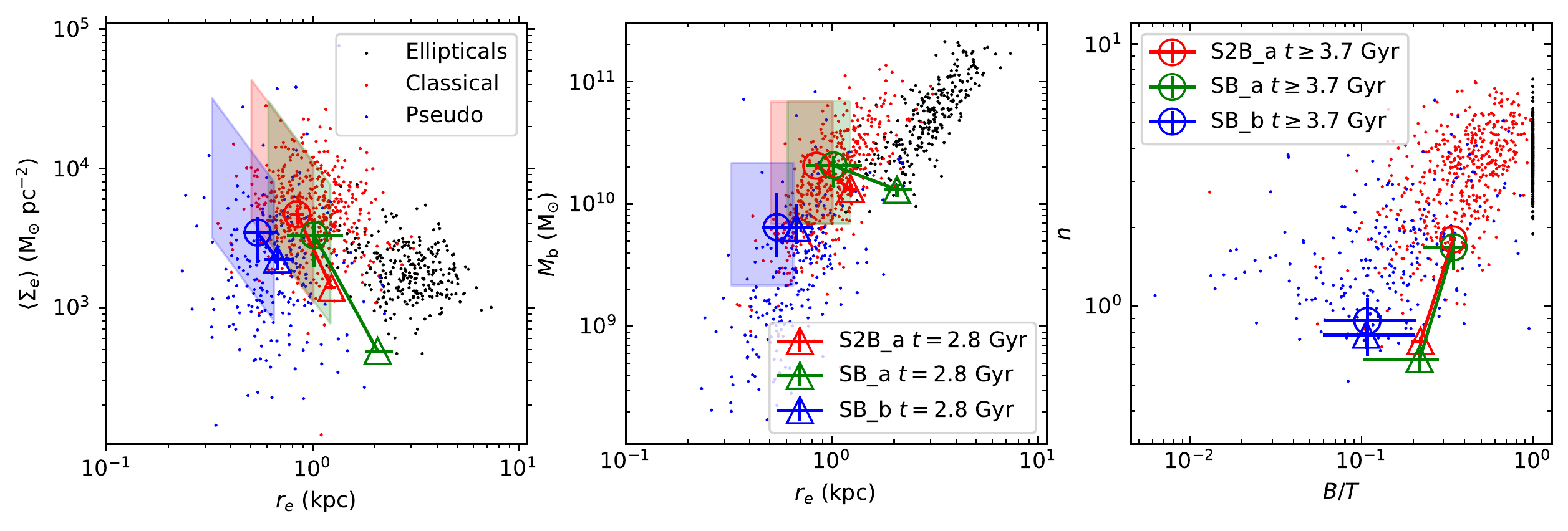}
    \caption{Scaling relations: (left) average surface mass density within the bulge effective radius $\langle\Sigma_e\rangle$ vs. the effective radius $r_e$; (middle) bulge stellar mass $M_{\rm b}$ vs. $r_e$, and (right) S\'ersic index $n$ vs. bulge-to-total ratio $B/T$. The black, red, and blue points represent ellipticals, classical bulges, and pseudo bulges, respectively, adopted from \citet{gadotti2009structural}. The results of our models are overlaid as open triangles (Phase 1, $t=2.8$ Gyr) and circles (Phase 3, $t\geq$ 3.7 Gyr) connected by solid lines. We average the results obtained from the decompositions using two bar functions and three inclination angles. Decomposition results from five snapshots of $t=3.7-7.4$ Gyr are used for the cases of Phase 3. The error bars correspond to upper and lower limits. The shaded regions in the left and middle panels show the regions coverd by reasonable scalings of the parameters of Phase 3 (see text).}
    \label{fig:scaling_relation}
\end{figure*}
\begin{figure*}[htb]
    \centering
    \includegraphics[width=0.7\linewidth]{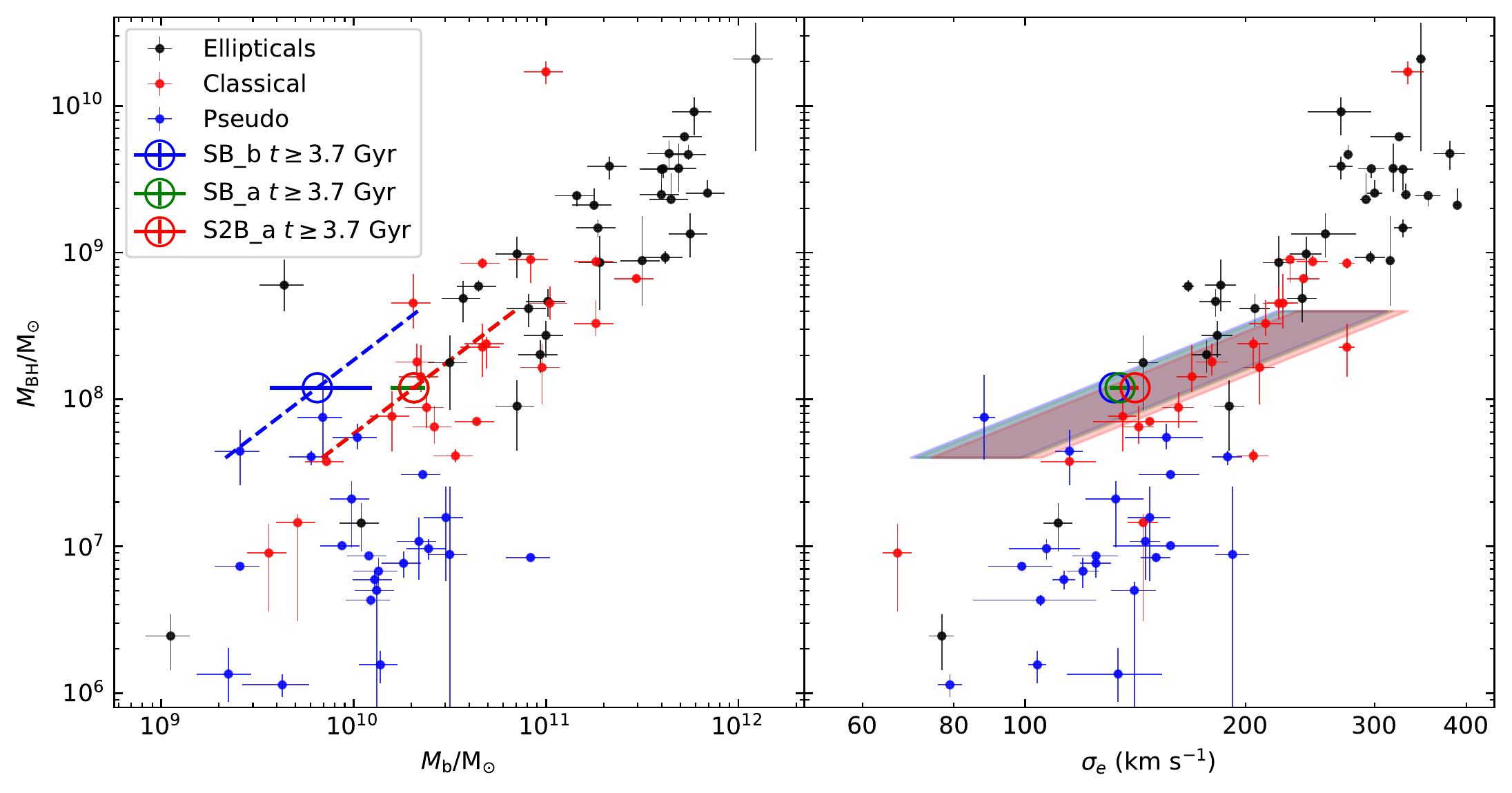}
    \caption{(Left) $M_{\rm BH} - M_{\rm b}$ and (right) $M_{\rm BH} - \sigma_e$ relation of the models, where $M_{\rm b}$ is the bulge stellar mass and $\sigma_e$ is the mean velocity dispersion within $r_e$. The parameters were obtained in the same manner as \reffig{fig:scaling_relation}. The data for the ellipticals, classical and pseudo bulges are from \citet{kormendy2013coevolution}. In the left panel, the mass ratio is constant along each dashed line. The red dashed line marks the constant mass ratio $M_{\rm BH} / M_{\rm b} \approx 0.006$. The shaded regions in the right panel show the reasonable scaling regions of the parameters (see text). Note that in left panel, SB\_a (green circle) is almost completely covered by S2B\_a (red circle).}
    \label{fig:BH-Bulge}
\end{figure*}

Pseudo bulges are distinct from classical bulge \citep[reviewed by][]{kormendy2004secular}. Classical bulges are thought to be spheroidal systems that form in violent, dissipative processes, such as gas-rich mergers \citep[e.g.,][]{Toomre1977merger, Aguerri2001growth}, while pseudo bulges likely originate from slow internal processes \citep{kormendy2011supermassive}. A widely accepted argument for advancing this theory is that classical bulges follow the same fundamental plane as elliptical galaxies \citep{kormendy2009structure, fisher2010bulges, kormendy2011supermassive}, whereas pseudo bulges, including boxy/peanut bulges, are generally offset from this relation.

We examine our simulated bulges on the three scaling relations that have been commonly used to classify bulges (\reffig{fig:scaling_relation}), overlaid on the data adapted from \citet{gadotti2009structural}. The observed bulges are classified according to the \citet{Kormendy1977ApJ...218..333K} relation of \citet[][their Figure 8]{gadotti2009structural}. The quantity $\langle\Sigma_e\rangle$ is the average stellar mass surface density within the half-mass radius $r_e$, derived using the average surface luminosity and mass-to-light ratio from \citet{gadotti2009structural}. We focus on the main group, as the bulges of the control group only evolve mildly at $t\geq 3.7$ Gyr. The simulation output for the two time steps are connected by a solid line. It is worth emphasizing that the absolute position of the model galaxies on the scaling relations are determined by an arbitrary scaling factor, and the most physically meaningful comparison is their {\it relative}\ evolution. The unfilled symbols indicate the results assuming a fiducial scaling of $M_{\star}=6\times10^{10}\, {\rm M_\odot}$ and $h_R=2.5\,{\rm kpc}$. To study the effect of unit scaling, we adjust the simulations to a reasonable mass range of $M_\star = 2\times10^{10}-2\times10^{11}\, {\rm M_\odot}$. During the steady stage (Phase 3), the scale length of the disk is about $2.0 \ h_R$. The disk scale length in observations \citep{Fathi2010scalelength} is $\sim 5$ kpc for galaxies in the same stellar mass range. We vary the length unit from 1.5 to 3 kpc, which covers the typical range of disk scale length for galaxies of this mass range. The results are shown by the shaded regions. At $t=2.8$ Gyr (open triangles), the model bulges have diverse properties. All the bulges become more massive and compact at $t\geq 3.7$ Gyr (open circles), moving upper-left on the $\langle\Sigma_e\rangle-r_e$ and $M_{\rm b}-r_e$ diagrams. The bulge in model SB\_b is relatively diffuse and less massive, thus falling among the pseudo bulge class. In comparison, at $t\geq 3.7$ Gyr the bulges in models S2B\_a and SB\_a are as massive and compact as observed classical bulges. The same trend holds for the relation between $n$ and $B/T$. Therefore, the growth of bulges driven by the destruction of short bars may significantly blur the difference between pseudo and classical bulges. Some relatively less massive and compact classical bulges may form via secular evolution due to the destruction of short bars. 

\reffig{fig:BH-Bulge} (left) shows the correlation between BH mass and bulge stellar mass, using as comparison the data assembled in \citet{kormendy2013coevolution}. The unfilled circles mark the results assuming the fiducial scaling for $t\geq3.7$ Gyr, after the BH has grown. The dashed lines and the shaded regions are obtained by the same method used in \reffig{fig:scaling_relation}.
Our simulations suggest that the mass ratio between BHs and bulges is constant at $M_{\rm BH} / M_{\rm b} \approx 0.006$ (red dashed line in the left panel). This is consistent with the median mass ratio observed in classical bulges. This constant $M_{\rm BH} / M_{\rm b}$ results from the nearly constant bulge-to-total mass ratio ($M_{\rm b} /M_\star\approx 0.35$) obtained in the simulations and from our imposing a maximum BH mass allowed for secular processes, as suggested by \citet{Du2017black}. Model SB\_b is clearly offset from the $M_{\rm BH}-M_{\rm b}$ relation because of the usage of the maximum BH mass. A galaxy hosting a single large-scale bar may not form BHs as massive as those in galaxies with short bars, and it may fall below the scaling relation of classical bulges and be more consistent with the pseudo bulges. Pseudo bulges may evolve toward the same scaling relation as classical bulges via the mechanism of short bar-mediated BH growth suggested in this study. 
 
Similarly, \reffig{fig:BH-Bulge} (right) plots the $M_{\rm BH}-\sigma_e$ relation, where $\sigma_e$ is the average velocity dispersion of the bulge measured within $r_e$. The shaded band corresponds to varying the length unit between 1.5 and 3 kpc. The resulting slope of $\sim 2$ is clearly too shallow to match the observed slope of 4.4 \citep{kormendy2013coevolution}. However, our present very limited set of models cannot possibly be expected to produce a realistic match to the observations. They only suffice to demonstrate that our simulated bulges bear a close resemblance to classical bulges. We note that all three models have similar values of $\sigma_e$, and $\sigma_e$ remains nearly the same during the growth of the BH, even though the bulge mass increases significantly and the bar structures evolve significantly.
In our models $\sigma_e$ may be largely determined by the total stellar mass of the system. In real galaxies or more realistic simulations, $\sigma_e$ would be reduced by the formation of new stars, which is not implemented in our 
current treatment. 

\begin{figure}[htb]
    \centering
    \includegraphics[width=0.9\linewidth]{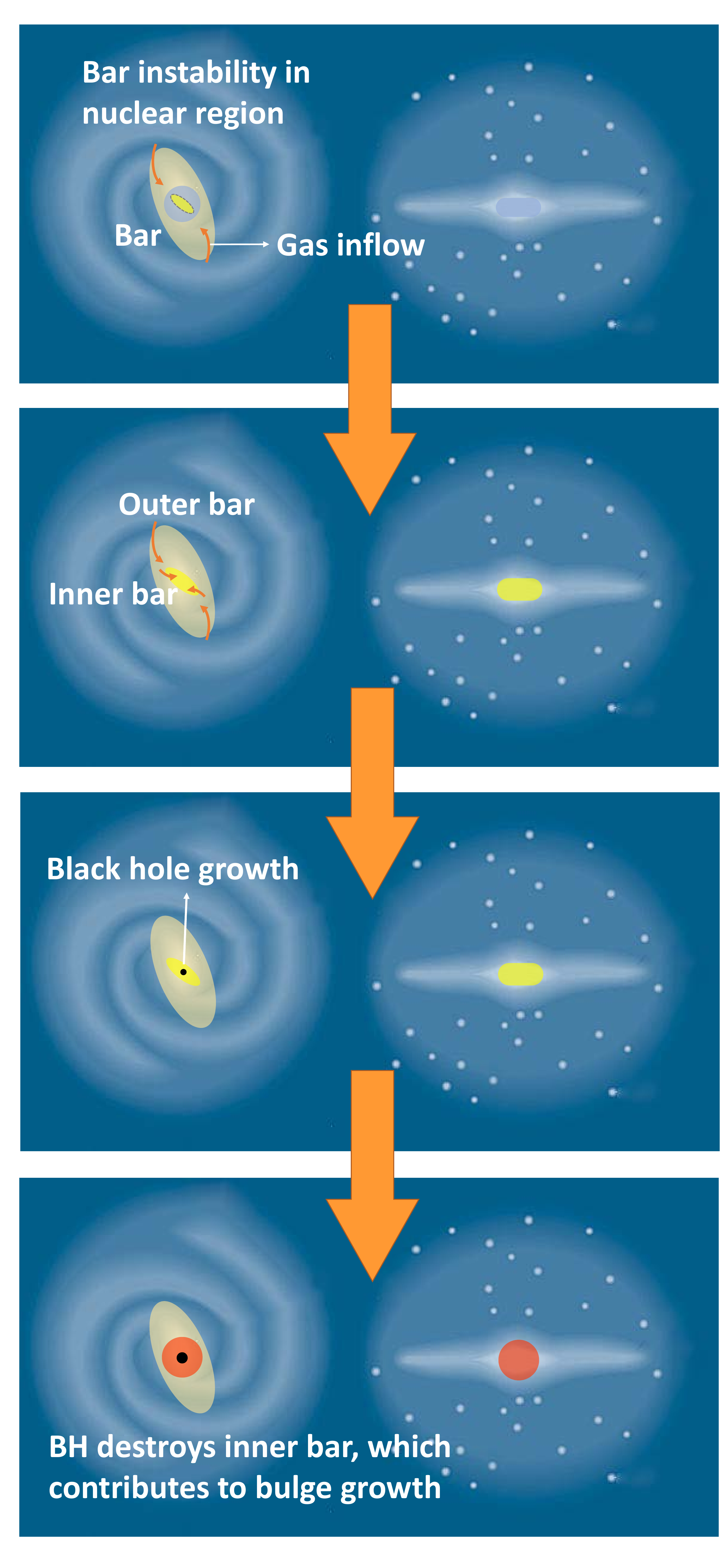}
    \caption{Schematic illustration of the scenario proposed in \refsec{sec:picture} on the coevolution between BHs and bulges due to the destruction of short bars. 
    }
    \label{fig:picture}
\end{figure}

\section{Discussion: A channel of forming classical bulges from disks/pseudo bulges}
\label{sec:picture}

We provide a potential evolutionary channel that allows disks to evolve into classical-like bulges without mergers. This scenario is illustrated in \reffig{fig:picture}. \citet{du2015forming} showed that a short (inner) bar forms spontaneously when a dynamically cool, inner disk (grey circle in the top panel) exists. Such an inner disk may be built up gradually by gas funneled inward due to perturbations from a lage-scale bars \citep{Athanassoula2005nature} or tidal forces \citep{Hernquist1989Tidal}. Short (inner) bars have been hypothesized to be an important mechanism for driving gas inflows on $<100$ pc scales, efficiently feeding BHs \citep[e.g.,][second panel of \reffig{fig:picture}]{Shlosman1989Bars}. \citet{Du2017black} suggested that the maximum mass a BH can reach via secular evolution is about $0.1\%$ of the total stellar mass, because a BH more massive that this threshold would destroy the inner bar. According to the result of this study, the remnants of short bars have properties similar to those of classical bulges. Therefore, an inner cold disk/bar can be transformed into a classical bulge under the influence of a BH. Classical bulges forming by this mechanism are likely to be younger and more metal-rich than typical classical bulges. It is interesting to note that classical bulges exhibit a clear bimodal distribution in age (Figure 9 of \citealp{gadotti2009structural}; see also \citealp{Allen2006Millennium}). Are the classical bulges with young stellar populations formed from the destruction of short bars? 

It is worth mentioning that the short bars invoked in our scenario need not be as stable and long-lived as those of our models. Many $N$-body+gas simulations generate gas-rich, short-lived nuclear bars \citep[e.g.,][]{Friedli&Martinet1993, Combes1994gas, Shlosman&Heller2002nested, Englmaier&Shlosman2004dynamical, Wozniak2015double}. Bulges may grow gradually by destroying recurring weak nuclear bars.

Scaling relations offer a useful, practical framework for constraining the formation and evolution of galaxies. While classical and pseudo bulges occupy statistically different loci in the mean (e.g., on the Kormendy relation; see \citealp{gadotti2009structural}; \citealp{Gao2019demographics}), no clear boundary separates the two types. We propose that galaxies with inner short bars offer a channel for pseudo bulges to migrate into the territory of classical bulges. Classical bulges, therefore, may be a mixed-bag; they are not all simply little ellipticals surrounded by disks. \citet{Breda&Papaderos2018continuous} also argue that secular processes in disks produce a continuum of bulge properties. The bulge formed by this new channel is not included in the bulge dichotomy proposed by \citet{kormendy2004secular}. Its moderate S\'ersic index is likely to blur the difference between a real classical bulge and pseudo bulge. Because of this, the S\'ersic index is not a clean discriminant between classical and pseudo bulges \citep{Gao2019classification}.

\section{Summary} \label{sec:sum}


Short ($\lesssim 1$ kpc) (inner) bars are an important mechanism for driving gas inflows to feed central black holes. Our previous work \citep{Du2017black} has shown that a black hole of mass $\sim 0.1\%$ of the total stellar mass of the galaxy can destroy the inner bar, thereby self-limiting the growth of the black hole. This paper examines in detail the consequences of this scenario for the central stellar distribution of the host galaxy. What is the fate of the remnant of the destroyed inner bar? We demonstrate that an initially boxy/peanut-shaped bulge with S\'ersic index $n \lesssim 1$ gets transformed into a more massive, compact ($n\approx 2$), isotropic, slowly rotating spheroid that bears many of the characteristics of a classical bulge and in terms of their location on bulge scaling relations. We propose a new channel for forming classical bulges from the destruction of short bars formed from nuclear disks.

\section*{Acknowledgement}

This work was supported by the National Science Foundation of China (11721303) and the National Key R\&D Program of China (2016YFA0400702). We thank Peter Erwin for the constructive discussion. M.D. is supported by the grants “National Postdoctoral Program for Innovative Talents” (No. 8201400810) and “Postdoctoral Science Foundation of China” (No. 8201400927) from the China Postdoctoral Science Foundation. V.P.D. was supported by STFC Consolidated grant ST/R000786/1. V.P.D. acknowledges support from the Kavli Visiting Scholars Program for a visit to the KIAA during this work. We utilized the High-performance Computing Platform of Peking University.

\bibliography{ref} 

\begin{thebibliography}{}
\expandafter\ifx\csname natexlab\endcsname\relax\def\natexlab#1{#1}\fi

\bibitem[{{Aguerri} {et~al.}(2001){Aguerri}, {Balcells}, \&
  {Peletier}}]{Aguerri2001growth}
{Aguerri}, J.~A.~L., {Balcells}, M., \& {Peletier}, R.~F. 2001, \aap, 367, 428

\bibitem[{{Allen} {et~al.}(2006){Allen}, {Driver}, {Graham}, {Cameron},
  {Liske}, \& {de Propris}}]{Allen2006Millennium}
{Allen}, P.~D., {Driver}, S.~P., {Graham}, A.~W., {et~al.} 2006, \mnras, 371, 2

\bibitem[{Athanassoula(1992)}]{Athanassoula1992morphology}
Athanassoula, E. 1992, \mnras, 259, 328

\bibitem[{{Athanassoula}(2004)}]{Athanassoula2004bars}
{Athanassoula}, E. 2004, in IAU Symposium, Vol. 220, Dark Matter in Galaxies,
  ed. S.~{Ryder}, D.~{Pisano}, M.~{Walker}, \& K.~{Freeman}, 255

\bibitem[{{Athanassoula}(2005)}]{Athanassoula2005nature}
{Athanassoula}, E. 2005, \mnras, 358, 1477

\bibitem[{Athanassoula {et~al.}(2005)Athanassoula, Lambert, \&
  Dehnen}]{Athanassoula2005bars}
Athanassoula, E., Lambert, J.~C., \& Dehnen, W. 2005, \mnras, 363, 496

\bibitem[{{Barnes} \& {Hernquist}(1996)}]{Barnes&Hernquist1996}
{Barnes}, J.~E., \& {Hernquist}, L. 1996, \apj, 471, 115

\bibitem[{{Barth} {et~al.}(2004){Barth}, {Ho}, {Rutledge}, \&
  {Sargent}}]{Barth2004POX52}
{Barth}, A.~J., {Ho}, L.~C., {Rutledge}, R.~E., \& {Sargent}, W. L.~W. 2004,
  ApJ, 607, 90

\bibitem[{{Bell} {et~al.}(2017){Bell}, {Monachesi}, {Harmsen}, {de Jong},
  {Bailin}, {Radburn-Smith}, {D'Souza}, \& {Holwerda}}]{Bell2017galaxies}
{Bell}, E.~F., {Monachesi}, A., {Harmsen}, B., {et~al.} 2017, \apj, 837, L8

\bibitem[{{Bournaud} {et~al.}(2008){Bournaud}, {Duc}, \&
  {Emsellem}}]{Bournaud2008high}
{Bournaud}, F., {Duc}, P.~A., \& {Emsellem}, E. 2008, \mnras, 389, L8

\bibitem[{{Breda} \& {Papaderos}(2018)}]{Breda&Papaderos2018continuous}
{Breda}, I., \& {Papaderos}, P. 2018, \aap, 614, A48

\bibitem[{{Cisternas} {et~al.}(2011){Cisternas}, {Jahnke}, {Inskip},
  {Kartaltepe}, {Koekemoer}, {Lisker}, {Robaina}, {Scodeggio}, {Sheth}, \&
  {Trump}}]{Cisternas2011bulk}
{Cisternas}, M., {Jahnke}, K., {Inskip}, K.~J., {et~al.} 2011, \apj, 726, 57

\bibitem[{{Clarke} {et~al.}(2019){Clarke}, {Debattista}, {Nidever}, {Loebman},
  {Simons}, {Kassin}, {Du}, {Ness}, {Fisher}, {Quinn}, {Wadsley}, {Freeman}, \&
  {Popescu}}]{Clarke&Debattista2019}
{Clarke}, A.~J., {Debattista}, V.~P., {Nidever}, D.~L., {et~al.} 2019, \mnras,
  484, 3476

\bibitem[{{Combes}(1994)}]{Combes1994gas}
{Combes}, F. 1994, in Mass-Transfer Induced Activity in Galaxies, ed.
  I.~{Shlosman}, 170

\bibitem[{{de Lorenzo-C{\'a}ceres} {et~al.}(2008){de Lorenzo-C{\'a}ceres},
  {Falc{\'o}n-Barroso}, {Vazdekis}, \&
  {Mart{\'\i}nez-Valpuesta}}]{deLorenzo-Caceres2008}
{de Lorenzo-C{\'a}ceres}, A., {Falc{\'o}n-Barroso}, J., {Vazdekis}, A., \&
  {Mart{\'\i}nez-Valpuesta}, I. 2008, \apjl, 684, L83

\bibitem[{{Debattista} {et~al.}(2004){Debattista}, {Carollo}, {Mayer}, \&
  {Moore}}]{Debattista2004bulge}
{Debattista}, V.~P., {Carollo}, C.~M., {Mayer}, L., \& {Moore}, B. 2004, \apj,
  604, L93

\bibitem[{{Debattista} {et~al.}(2002){Debattista}, {Corsini}, \&
  {Aguerri}}]{Debattista2002fast}
{Debattista}, V.~P., {Corsini}, E.~M., \& {Aguerri}, J.~A.~L. 2002, \mnras,
  332, 65

\bibitem[{{Debattista} {et~al.}(2006){Debattista}, {Mayer}, {Carollo}, {Moore},
  {Wadsley}, \& {Quinn}}]{Debattista2006secular}
{Debattista}, V.~P., {Mayer}, L., {Carollo}, C.~M., {et~al.} 2006, \apj, 645,
  209

\bibitem[{{Debattista} \& {Shen}(2007)}]{Debattista2007}
{Debattista}, V.~P., \& {Shen}, J. 2007, \apjl, 654, L127

\bibitem[{{Dekel} {et~al.}(2009){Dekel}, {Sari}, \&
  {Ceverino}}]{Dekel2009formation}
{Dekel}, A., {Sari}, R., \& {Ceverino}, D. 2009, \apj, 703, 785

\bibitem[{Du {et~al.}(2016)Du, Debattista, Shen, \&
  Cappellari}]{du2016kinematic}
Du, M., Debattista, V.~P., Shen, J., \& Cappellari, M. 2016, ApJ, 828, 14

\bibitem[{Du {et~al.}(2017)Du, Debattista, Shen, Ho, \& Erwin}]{Du2017black}
Du, M., Debattista, V.~P., Shen, J., Ho, L.~C., \& Erwin, P. 2017, ApJL, 844,
  L15

\bibitem[{Du {et~al.}(2015)Du, Shen, \& Debattista}]{du2015forming}
Du, M., Shen, J., \& Debattista, V.~P. 2015, ApJ, 804, 139

\bibitem[{{Du} {et~al.}(2017){Du}, {Shen}, {Debattista}, \& {de
  Lorenzo-C{\'a}ceres}}]{du2017sigma}
{Du}, M., {Shen}, J., {Debattista}, V.~P., \& {de Lorenzo-C{\'a}ceres}, A.
  2017, \apj, 836, 181

\bibitem[{{Elmegreen} {et~al.}(2008){Elmegreen}, {Bournaud}, \&
  {Elmegreen}}]{Elmegreen2008bulge}
{Elmegreen}, B.~G., {Bournaud}, F., \& {Elmegreen}, D.~M. 2008, \apj, 688, 67

\bibitem[{{Elmegreen} \& {Elmegreen}(2005)}]{Elmegreen&Elmegreen2005stellar}
{Elmegreen}, B.~G., \& {Elmegreen}, D.~M. 2005, \apj, 627, 632

\bibitem[{{Englmaier} \& {Shlosman}(2004)}]{Englmaier&Shlosman2004dynamical}
{Englmaier}, P., \& {Shlosman}, I. 2004, \apj, 617, L115

\bibitem[{Erwin(2004)}]{erwin2004double}
Erwin, P. 2004, \aap, 415, 941

\bibitem[{Erwin \& Sparke(2002)}]{erwin2002double}
Erwin, P., \& Sparke, L.~S. 2002, \aj, 124, 65

\bibitem[{{Fathi} {et~al.}(2010){Fathi}, {Allen}, {Boch}, {Hatziminaoglou}, \&
  {Peletier}}]{Fathi2010scalelength}
{Fathi}, K., {Allen}, M., {Boch}, T., {Hatziminaoglou}, E., \& {Peletier},
  R.~F. 2010, \mnras, 406, 1595

\bibitem[{{Ferrarese} \& {Merritt}(2000)}]{Ferrarese&Merritt2000}
{Ferrarese}, L., \& {Merritt}, D. 2000, \apjl, 539, L9

\bibitem[{{Filippenko} \& {Ho}(2003)}]{Filippenko&Ho2003lowmass}
{Filippenko}, A.~V., \& {Ho}, L.~C. 2003, ApJ, 588, L13

\bibitem[{{Fisher} \& {Drory}(2008)}]{Fisher&Drory2008AJ....136..773F}
{Fisher}, D.~B., \& {Drory}, N. 2008, \aj, 136, 773

\bibitem[{Fisher \& Drory(2010)}]{fisher2010bulges}
Fisher, D.~B., \& Drory, N. 2010, ApJ, 716, 942

\bibitem[{{Friedli} \& {Martinet}(1993)}]{Friedli&Martinet1993}
{Friedli}, D., \& {Martinet}, L. 1993, \aap, 277, 27

\bibitem[{Gadotti(2009)}]{gadotti2009structural}
Gadotti, D.~A. 2009, \mnras, 393, 1531

\bibitem[{{Gao} {et~al.}(2019{\natexlab{a}}){Gao}, {Ho}, {Barth}, \&
  {Li}}]{Gao2019classification}
{Gao}, H., {Ho}, L.~C., {Barth}, A.~J., \& {Li}, Z.-Y. 2019{\natexlab{a}},
  \apjs, submitted

\bibitem[{{Gao} {et~al.}(2019{\natexlab{b}}){Gao}, {Ho}, {Barth}, \&
  {Li}}]{Gao2019demographics}
---. 2019{\natexlab{b}}, \apjs, 244, 34

\bibitem[{Gebhardt {et~al.}(2000)Gebhardt, Bender, Bower, Dressler, Faber,
  Filippenko, Green, Grillmair, Ho, Kormendy,
  {et~al.}}]{gebhardt2000relationship}
Gebhardt, K., Bender, R., Bower, G., {et~al.} 2000, ApJL, 539, L13

\bibitem[{{Genzel} {et~al.}(2008){Genzel}, {Burkert}, {Bouch{\'e}}, {Cresci},
  {F{\"o}rster Schreiber}, {Shapley}, {Shapiro}, {Tacconi}, {Buschkamp}, \&
  {Cimatti}}]{Genzel2008ring}
{Genzel}, R., {Burkert}, A., {Bouch{\'e}}, N., {et~al.} 2008, \apj, 687, 59

\bibitem[{Gerhard \& Binney(1985)}]{gerhard1985triaxial}
Gerhard, O.~O., \& Binney, J. 1985, \mnras, 216, 467

\bibitem[{{Greene} {et~al.}(2008){Greene}, {Ho}, \& {Barth}}]{Greene2008black}
{Greene}, J.~E., {Ho}, L.~C., \& {Barth}, A.~J. 2008, ApJ, 688, 159

\bibitem[{{Greene} {et~al.}(2019){Greene}, {Strader}, \&
  {Ho}}]{Greene2019review}
{Greene}, J.~E., {Strader}, J., \& {Ho}, L.~C. 2019, \araa, in press

\bibitem[{G{\"u}ltekin {et~al.}(2009)G{\"u}ltekin, Richstone, Gebhardt, Lauer,
  Tremaine, Aller, Bender, Dressler, Faber, Filippenko,
  {et~al.}}]{gultekin2009m}
G{\"u}ltekin, K., Richstone, D.~O., Gebhardt, K., {et~al.} 2009, ApJ, 698, 198

\bibitem[{{H{\"a}ring} \& {Rix}(2004)}]{Haring&Rix2004black}
{H{\"a}ring}, N., \& {Rix}, H.-W. 2004, ApJ, 604, L89

\bibitem[{{Hasan} \& {Norman}(1990)}]{Hasan1990chaotic}
{Hasan}, H., \& {Norman}, C. 1990, \apj, 361, 69

\bibitem[{{Hernquist}(1989)}]{Hernquist1989Tidal}
{Hernquist}, L. 1989, \nat, 340, 687

\bibitem[{{Ho}(2008)}]{Ho2008nuclear}
{Ho}, L.~C. 2008, \araa, 46, 475

\bibitem[{{Hopkins} {et~al.}(2009){Hopkins}, {Cox}, {Younger}, \&
  {Hernquist}}]{Hopkins2009a}
{Hopkins}, P.~F., {Cox}, T.~J., {Younger}, J.~D., \& {Hernquist}, L. 2009,
  \apj, 691, 1168

\bibitem[{{Hopkins} \& {Quataert}(2010)}]{Hopkins&Quataert2010massive}
{Hopkins}, P.~F., \& {Quataert}, E. 2010, \mnras, 407, 1529

\bibitem[{{Hozumi}(2012)}]{Hozumi2012destructible}
{Hozumi}, S. 2012, \pasj, 64, 5

\bibitem[{{Inoue} \& {Saitoh}(2012)}]{Inoue&Saitoh2012nature}
{Inoue}, S., \& {Saitoh}, T.~R. 2012, \mnras, 422, 1902

\bibitem[{{Jiang} {et~al.}(2011){Jiang}, {Greene}, {Ho}, {Xiao}, \&
  {Barth}}]{Jiang2011host}
{Jiang}, Y.-F., {Greene}, J.~E., {Ho}, L.~C., {Xiao}, T., \& {Barth}, A.~J.
  2011, ApJ, 742, 68

\bibitem[{{Kim} {et~al.}(2017){Kim}, {Ho}, {Peng}, {Barth}, \&
  {Im}}]{Kim2017sellar}
{Kim}, M., {Ho}, L.~C., {Peng}, C.~Y., {Barth}, A.~J., \& {Im}, M. 2017, \apjs,
  232, 21

\bibitem[{{Kim} {et~al.}(2012){Kim}, {Seo}, \& {Kim}}]{Kim2012gaseous}
{Kim}, W.-T., {Seo}, W.-Y., \& {Kim}, Y. 2012, \apj, 758, 14

\bibitem[{{Kormendy}(1977)}]{Kormendy1977ApJ...218..333K}
{Kormendy}, J. 1977, ApJ, 218, 333

\bibitem[{Kormendy {et~al.}(2011)Kormendy, Bender, \&
  Cornell}]{kormendy2011supermassive}
Kormendy, J., Bender, R., \& Cornell, M. 2011, Nature, 469, 374

\bibitem[{Kormendy {et~al.}(2009)Kormendy, Fisher, Cornell, \&
  Bender}]{kormendy2009structure}
Kormendy, J., Fisher, D.~B., Cornell, M.~E., \& Bender, R. 2009, \apjs, 182,
  216

\bibitem[{Kormendy \& Ho(2013)}]{kormendy2013coevolution}
Kormendy, J., \& Ho, L.~C. 2013, \araa, 51, 511

\bibitem[{Kormendy \& Kennicutt(2004)}]{kormendy2004secular}
Kormendy, J., \& Kennicutt, R.~C. 2004, \araa, 42, 603

\bibitem[{Laine {et~al.}(2002)Laine, Shlosman, Knapen, \&
  Peletier}]{laine2002nested}
Laine, S., Shlosman, I., Knapen, J.~H., \& Peletier, R.~F. 2002, ApJ, 567, 97

\bibitem[{{Li} {et~al.}(2015){Li}, {Shen}, \& {Kim}}]{LiZhi2015hydrodynamical}
{Li}, Z., {Shen}, J., \& {Kim}, W.-T. 2015, \apj, 806, 150

\bibitem[{Noguchi(1998)}]{noguchi1998clumpy}
Noguchi, M. 1998, Nature, 392, 253

\bibitem[{Noguchi(1999)}]{noguchi1999early}
---. 1999, ApJ, 514, 77

\bibitem[{{Okamoto}(2013)}]{Okamoto2013origin}
{Okamoto}, T. 2013, \mnras, 428, 718

\bibitem[{{Park} {et~al.}(2019){Park}, {Yi}, {Dubois}, {Pichon}, {Kimm},
  {Devriendt}, {Choi}, {Volonteri}, {Kaviraj}, \& {Peirani}}]{Park2019horizon}
{Park}, M.-J., {Yi}, S.~K., {Dubois}, Y., {et~al.} 2019, \apj, 883, 25

\bibitem[{Peng {et~al.}(2002)Peng, Ho, Impey, \& Rix}]{peng2002detailed}
Peng, C.~Y., Ho, L.~C., Impey, C.~D., \& Rix, H.-W. 2002, \aj, 124, 266

\bibitem[{Peng {et~al.}(2010)Peng, Ho, Impey, \& Rix}]{peng2010detailed}
---. 2010, \aj, 139, 2097

\bibitem[{Pfenniger \& Norman(1990)}]{pfenniger1990dissipation}
Pfenniger, D., \& Norman, C. 1990, ApJ, 363, 391

\bibitem[{Raha {et~al.}(1991)Raha, Sellwood, James, \&
  Kahn}]{raha1991dynamical}
Raha, N., Sellwood, J., James, R., \& Kahn, F. 1991, Nature, 352, 411

\bibitem[{{Reines} \& {Volonteri}(2015)}]{Reines2015relations}
{Reines}, A.~E., \& {Volonteri}, M. 2015, \apj, 813, 82

\bibitem[{{Sales} {et~al.}(2012){Sales}, {Navarro}, {Theuns}, {Schaye},
  {White}, {Frenk}, {Crain}, \& {Dalla Vecchia}}]{Sales2012origin}
{Sales}, L.~V., {Navarro}, J.~F., {Theuns}, T., {et~al.} 2012, \mnras, 423,
  1544

\bibitem[{{Scannapieco} {et~al.}(2009){Scannapieco}, {White}, {Springel}, \&
  {Tissera}}]{Scannapieco2009formation}
{Scannapieco}, C., {White}, S. D.~M., {Springel}, V., \& {Tissera}, P.~B. 2009,
  \mnras, 396, 696

\bibitem[{Sellwood(2014)}]{sellwood2014galaxy}
Sellwood, J. 2014, arXiv preprint arXiv:1406.6606

\bibitem[{{Shen} \& {Sellwood}(2004)}]{Shen2004destruction}
{Shen}, J., \& {Sellwood}, J.~A. 2004, \apj, 604, 614

\bibitem[{{Shlosman} {et~al.}(1989){Shlosman}, {Frank}, \&
  {Begelman}}]{Shlosman1989Bars}
{Shlosman}, I., {Frank}, J., \& {Begelman}, M.~C. 1989, \nat, 338, 45

\bibitem[{{Shlosman} \& {Heller}(2002)}]{Shlosman&Heller2002nested}
{Shlosman}, I., \& {Heller}, C.~H. 2002, \apj, 565, 921

\bibitem[{{Toomre}(1977)}]{Toomre1977merger}
{Toomre}, A. 1977, in Evolution of Galaxies and Stellar Populations, ed. B.~M.
  {Tinsley} \& D.~C. {Larson}, Richard B.~Gehret, 401

\bibitem[{{Tremaine} {et~al.}(2002){Tremaine}, {Gebhardt}, {Bender}, {Bower},
  {Dressler}, {Faber}, {Filippenko}, {Green}, {Grillmair}, {Ho}, {Kormendy},
  {Lauer}, {Magorrian}, {Pinkney}, \& {Richstone}}]{Tremaine2002slope}
{Tremaine}, S., {Gebhardt}, K., {Bender}, R., {et~al.} 2002, ApJ, 574, 740

\bibitem[{{Vandenberg} {et~al.}(1996){Vandenberg}, {Bolte}, \&
  {Stetson}}]{Vandenberg1996age}
{Vandenberg}, D.~A., {Bolte}, M., \& {Stetson}, P.~B. 1996, \araa, 34, 461

\bibitem[{{Wang} {et~al.}(2019){Wang}, {Obreschkow}, {Lagos}, {Sweet},
  {Fisher}, {Glazebrook}, {Macci{\`o}}, {Dutton}, \& {Kang}}]{Wang2019angular}
{Wang}, L., {Obreschkow}, D., {Lagos}, C. d.~P., {et~al.} 2019, \mnras, 482,
  5477

\bibitem[{{Wozniak}(2015)}]{Wozniak2015double}
{Wozniak}, H. 2015, \aap, 575, A7

\bibitem[{{Wu} {et~al.}(2016){Wu}, {Pfenniger}, \&
  {Taam}}]{Wu2016time-dependent}
{Wu}, Y.-T., {Pfenniger}, D., \& {Taam}, R.~E. 2016, \apj, 830, 111

\bibitem[{{Zhao} {et~al.}(2019){Zhao}, {Ho}, {Zhao}, {Shangguan}, \&
  {Kim}}]{ZhaoDongyao2019role}
{Zhao}, D., {Ho}, L.~C., {Zhao}, Y., {Shangguan}, J., \& {Kim}, M. 2019, \apj,
  877, 52

\bibitem[{{Zolotov} {et~al.}(2015){Zolotov}, {Dekel}, {Mandelker}, {Tweed},
  {Inoue}, {DeGraf}, {Ceverino}, {Primack}, {Barro}, \& {Faber}}]{Zolotov2015}
{Zolotov}, A., {Dekel}, A., {Mandelker}, N., {et~al.} 2015, \mnras, 450, 2327

\end{thebibliography}


\end{document}